\documentclass[12pt,onecolumn]{IEEEtran} 
\usepackage{graphicx}
\usepackage{array} 
\usepackage{verbatim}
\usepackage{latexsym}     
\usepackage{amssymb}
\usepackage{amsbsy}
\usepackage{amsmath}
\usepackage{algorithm}
\usepackage{algorithmic}
\usepackage{subfig}

\newcommand{\vect}[0]{\text{vect}}

\newcommand{\vectdiag}[0]{\text{vectdiag}}
\newcommand{\rank}[0]{\text{rank}}

\newcommand{\diag}[0]{\text{diag}}

\newcommand{\Var}[0]{\text{Var}}

\newcommand{\sigmah}[0]{\hat{\sigma}}

\newcommand{\bbeta}[0]{\boldsymbol{\beta}}

\newcommand{\bgamma}[0]{\boldsymbol{\gamma}}

\newcommand{\bdelta}[0]{\boldsymbol{\delta}}

\newcommand{\bzeta}[0]{\boldsymbol{\zeta}}

\newcommand{\btheta}[0]{\boldsymbol{\theta}}

\newcommand{\blambda}[0]{\boldsymbol{\lambda}}
\newcommand{\bsigma}[0]{\boldsymbol{\sigma}}
\newcommand{\bsigmah}[0]{\boldsymbol{\hat{\sigma}}}
\newcommand{\bsigmat}[0]{\boldsymbol{\tilde{\sigma}}}

\newcommand{\bSigmat}[0]{\boldsymbol{\tilde{\Sigma}}}

\newcommand{\bXi}[0]{\boldsymbol{\Xi}}

\newcommand{\bPsit}[0]{\boldsymbol{\tilde{\Psi}}}

\newcommand{\ba}[0]{\mathbf{a}}
\newcommand{\bA}[0]{\mathbf{A}}
\newcommand{\bb}[0]{\mathbf{b}}
\newcommand{\bB}[0]{\mathbf{B}}
\newcommand{\bc}[0]{\mathbf{c}}
\newcommand{\bC}[0]{\mathbf{C}}

\newcommand{\bD}[0]{\mathbf{D}}
\newcommand{\be}[0]{\mathbf{e}}

\newcommand{\bg}[0]{\mathbf{g}}
\newcommand{\bG}[0]{\mathbf{G}}
\newcommand{\bh}[0]{\mathbf{h}}
\newcommand{\bH}[0]{\mathbf{H}}

\newcommand{\bI}[0]{\mathbf{I}}

\newcommand{\bk}[0]{\mathbf{k}}
\newcommand{\bK}[0]{\mathbf{K}}

\newcommand{\bM}[0]{\mathbf{M}}
\newcommand{\bn}[0]{\mathbf{n}}

\newcommand{\bp}[0]{\mathbf{p}}

\newcommand{\bQ}[0]{\mathbf{Q}}
\newcommand{\br}[0]{\mathbf{r}}
\newcommand{\bR}[0]{\mathbf{R}}
\newcommand{\bs}[0]{\mathbf{s}}
\newcommand{\bS}[0]{\mathbf{S}}

\newcommand{\bu}[0]{\mathbf{u}}
\newcommand{\bU}[0]{\mathbf{U}}
\newcommand{\bv}[0]{\mathbf{v}}
\newcommand{\bV}[0]{\mathbf{V}}
\newcommand{\bw}[0]{\mathbf{w}}

\newcommand{\bx}[0]{\mathbf{x}}
\newcommand{\bX}[0]{\mathbf{X}}
\newcommand{\by}[0]{\mathbf{y}}

\newcommand{\bz}[0]{\mathbf{z}}

\newcommand{\EExp}[0]{\mathcal{E}}

\newcommand{\bAtt}[0]{\mathbf{\tilde{A}}}
\newcommand{\bat}[0]{\mathbf{\tilde{a}}}

\newcommand{\bxt}[0]{\mathbf{\tilde{x}}}

\newcommand{\brh}[0]{\mathbf{\hat{r}}}
\newcommand{\bRh}[0]{\mathbf{\hat{R}}}

\newcommand{\zeros}[0]{\mathbf{0}}
\newcommand{\ones}[0]{\mathbf{1}}

\newcommand{\MCA}[0]{\mathcal{A}}

\newcommand{\MCE}[0]{\mathcal{E}}

\newcommand{\MCF}[0]{\mathcal{F}}

\newcommand{\MCI}[0]{\mathcal{I}}

\newcommand{\MCL}[0]{\mathcal{L}}

\newcommand{\MCN}[0]{\mathcal{N}}

\newcommand{\MCU}[0]{\mathcal{U}}

\newcommand{\MCW}[0]{\mathcal{W}}

\newcommand{\MCX}[0]{\mathcal{X}}

\newcommand{\beq}{\begin{equation}}
\newcommand{\eeq}{\end{equation}}
\newcommand{\bea}{\begin{array}}
\newcommand{\ena}{\end{array}}
\newcommand{\vzero}{\mbox{${\mathbf 0}$}}

\newcommand{\bAt}[0]{\boldsymbol{\Psi}}

\newcommand{\MFDI}[0]{{\boldsymbol{\sigma}}_{\text{MF}}}
\newcommand{\MFDIind}[1]{\sigma_{\text{MF},#1}}

\newcommand{\SMFDI}[0]{\bsigmah_{\text{MF}}}
\newcommand{\SMFDIind}[1]{\sigmah_{\text{MF},#1}}

\newcommand{\SWLSDI}[0]{\bsigmah_{\text{WLS}}}

\newcommand{\MVDRDI}[0]{{\boldsymbol{\sigma}}_{\text{MVDR}}}
\newcommand{\MVDRDIind}[1]{\sigma_{\text{MVDR},#1}}

\newcommand{\UMVDRDI}[0]{\bsigmah_{\text{MVDR}}}
\newcommand{\UMVDRDIind}[1]{\sigmah_{\text{MVDR},#1}}

\newcommand{\WTDIind}[1]{\sigma_{\bw,#1}}

\newcommand{\HLS}[0]{\bH_{\text{LS}}}
\newcommand{\HWLS}[0]{\bH_{\text{WLS}}}

\newcommand{\PWLSsigma}[0]{\check{\bsigma}}

\title{Radio Astronomical Image Formation using Constrained Least Squares and Krylov Subspaces}
\author{A. Mouri Sardarabadi$^{1*}$, \and Amir Leshem$^2$, \and Alle-Jan van der Veen$^1$\thanks{This research was supported by NWO-TOP 2010, 614.00.005. The research of A. Leshem was supported by the Israeli Science foundation, grant 1240-2009. $^1$ Department of electrical engineering, Delft University of Technology. $^2$ Faculty of Engineering, Bar-Ilan University. $^*$Corresponding author, email: a.mourisardarabadi@tudelft.nl}}
\begin{document}
\maketitle
\begin{abstract}
Image formation for radio astronomy can be defined as estimating the
spatial power distribution of celestial sources over the sky, given an
array of antennas.  One of the challenges with image formation is that
the problem becomes ill-posed as the number of pixels becomes large.
The introduction of constraints that incorporate a-priori knowledge is
crucial.  In this paper we show that in addition to non-negativity, the
magnitude of each pixel in an image is also bounded from above.  Indeed,
the classical ``dirty image'' is an upper bound, but a much tighter
upper bound can be formed from the data using array processing
techniques.  This formulates image formation as a least squares
optimization problem with inequality constraints.  We propose to solve
this constrained least squares problem using active set techniques, and
the steps needed to implement it are described.  It is shown that the
least squares part of the problem can be efficiently implemented with
Krylov subspace based techniques, where the structure of the problem
allows massive parallelism and reduced storage needs.  The performance
of the algorithm is evaluated using simulations.
\end{abstract}

Image formation for radio astronomy can be defined as estimating the
spatial power distribution of celestial sources over the sky.  The data
model (``measurement equation'') is linear in the source powers, and the
resulting least squares problem has classically been implemented in two
steps: formation of a ``dirty image'', followed by a deconvolution step.
In this process, an implicit model assumption is made that the
number of sources is discrete, and subsequently the number of sources
has been replaced by the number of image pixels (assuming each pixel may
contain a source).

The deconvolution step becomes ill-conditioned if the number of pixels
is large \cite{wijnholds2008}.  
Alternatively, the directions of sources may be estimated
along with their powers, but this is a complex non-linear problem.
Classically, this has been implemented as an iterative subtraction
technique, wherein source directions are estimated from the dirty image,
and their contribution is subtracted from the data.  This mixed approach
is the essence of the CLEAN method proposed by H\"{o}gbom
\cite{hogbom74}, which was subsequently refined and extended in several
ways, leading to the widely used approaches described in
\cite{Cornwell2008,rao2009}.

    The conditioning of the image deconvolution step can be improved by
    incorporating side information such as non-negativity of the image
    \cite{briggs95}, source model structure beyond simple point sources
    (e.g., shapelets and wavelets \cite{reid2006}), and sparsity or $\ell_1$
    constraints on the image \cite{levanda08,wiaux09}.
    Beyond these, some fundamental approaches based on parameter estimation
    techniques have been proposed, such as the Least Squares Minimum
    Variance Imaging (LS-MVI) \cite{bendavid08} and maximum likelihood
    based techniques \cite{leshem2000a}.  Computational complexity is a
    concern and this has not been addressed in these approaches.

New radio telescopes such as the Low Frequency Array (LOFAR), the Allen
Telescope Array (ATA), Murchison Widefield Array (MWA) and the Long Wavelength Array (LWA) are composed
of many stations (each station made up of multiple antennas that are
combined using adaptive beamforming), and the increase in number of
antennas and stations continues in the design of the square kilometer
array (SKA).  These instruments have or will have a significantly
increased sensitivity and a larger field of view compared to traditional
telescopes, leading to many more sources that need to be taken into
account.  They also need to process larger bandwidths to reach this
sensitivity.  Besides the increased requirements on the performance of
imaging, the improved spatial resolution leads to an increasing number
of pixels in the image, and the development of computationally efficient
techniques is critical.

To benefit from the vast literature related to solving least
square problems, but also to gain from the non-linear processing
offered by standard deconvolution techniques, we
propose to reformulate the imaging problem as a parameter estimation
problem described by a weighted least squares optimization
problem with several constraints. 
The first is a non-negativity constraint, which would lead to the
non-negative least squares algorithm (NNLS) proposed in \cite{briggs95}.
But we show that the pixel values are also bounded from above.  A coarse
upper bound is provided by the classical dirty image, and a much tighter
bound is the ``minimum variance distortionless response'' (MVDR) dirty
image that was proposed in the context of radio astronomy in \cite{leshem2000a}.

We propose to solve the resulting constrained least squares problems using
an active set approach. 
This results in a computationally efficient imaging algorithm that is
closely related to existing non-linear sequential source estimation
techniques such as CLEAN with the benefit of accelerated convergence due
to tighter upper bounds on the power distribution over the complete
image.  Because the constraints are enforced over the entire image, this
eliminates the inclusion of negative flux sources and other anomalies
that appear in some existing sequential techniques.

To further reduce the computational complexity we show that the data
model has a Khatri-Rao structure.  This can be exploited to
significantly improve the data management and parallelism compared to
general implementations of least squares algorithms.

The structure of the paper is as follows. In Sec.\ \ref{sec:datam} we
describe the basic data model, and in Sec.\ \ref{sec:imagingp}
the image formation problem.  A constrained least squares problem is
formulated, using various power constraints that take the form of dirty
images. The solution of this problem using active set techniques in Sec.\
\ref{sec:COPAS} generalizes the classical CLEAN algorithm.
In Sec.\ \ref{sec:krylov} we discuss the efficient implementation of a key
step in the active set solution using Krylov subspaces.
We end up with some simulated experiments
demonstrating the advantages of the proposed technique and 
conclusions regarding future implementation.

\subsection*{Notation}
A boldface letter such as $\ba$ denotes a column vector, a boldface
capital letter such as $\bA$ denotes a matrix. We will frequently use
indexed matrices $\bA_k$ and let $\ba_{i,k}$ be the 
$i$th column of $\bA_k$, whereas $a_{i,k}$ is the $i$th element of
the vector $\ba_k$. 
$\bI$ is an identity matrix of appropriate size and $\bI_p$ is a
$p\times p$ identity matrix.

$(\cdot)^T$ is the transpose operator, $(\cdot)^*$ is the complex
conjugate operator, $(\cdot)^H$ is the Hermitian transpose,
$\|\cdot\|_F$ is the Frobenius norm of a matrix, $\|.\|$ is the two
norm of a vector and $\EExp\{\cdot\}$ is the expectation operator.

A calligraphic capital letter such as $\MCX$ represents a set of
indices, 

and $\ba_{\MCX}$ is a column vector constructed by stacking the
elements of $\ba$ that belong to $\MCX$.  The corresponding indices are
stored with the vector as well (similar to the storage of matlab ``sparse''
vectors).

$\vect(\cdot)$ stacks the columns of the argument matrix to form a
vector, $\vectdiag(\cdot)$ stacks the diagonal elements of the argument
matrix to form a vector, $\diag(\cdot)$ is a diagonal matrix with its
diagonal entries from the argument vector (if the argument is a matrix
$\diag(\cdot)=\diag(\vectdiag(\cdot))$).

Let $\otimes$ denote the Kronecker product, $\circ$ the Khatri-Rao
product (column-wise Kronecker product), and $\odot$ the Hadamard 
(element-wise) product.
The following properties are used throughout the paper (for matrices and
vectors with compatible dimensions):
\begin{align*}
(\bB^T \otimes \bA)\vect(\bX)&=\vect(\bA\bX\bB) \\
(\bB \otimes \bA)^H&=(\bB^H \otimes \bA^H)\\
(\bB \otimes \bA)^{-1}&=(\bB^{-1} \otimes \bA^{-1})\\
(\bB^T \circ \bA)\bx&=\vect(\bA\diag(\bx)\bB)\\
(\bB\bC \otimes \bA\bD)&=(\bB \otimes \bA)(\bC \otimes \bD)\\
(\bB\bC \circ \bA\bD)&=(\bB \otimes \bA)(\bC \circ \bD)\\
(\bB^H\bC \odot \bA^H\bD)&=(\bB \circ \bA)^H(\bC \circ \bD)\\
\vectdiag(\bA^H\bX\bA)&=(\bA^* \circ \bA)^H\vect(\bX)
\end{align*}

\section{Data Model}
\label{sec:datam}

We consider an instrument where $P$ receivers (stations or antennas) are observing the sky.
Assuming a discrete point source model, we let $Q$ denote the number of 
visible sources.
The received signals at the antennas are sampled and subsequently split
into narrow sub-bands.  For simplicity, we will consider only a single
sub-band in the rest of the paper.
Although the sources are considered stationary, because of the earth's
rotation the apparent position of the celestial sources will change with
time.  For this reason the data is split into short blocks or
``snapshots'' of $N$ samples, where the exact value of $N$ depends on
the resolution of the instrument.

We stack the output of the $P$ antennas at a single sub-band into a
vector $\by_k[n]$, where $n = 1, \cdots, N$ denotes the sample index,
and $k = 1, \cdots, K$ denotes the snapshot index.

The signals of the $q$th source arrive at the array with slight
delays for each antenna which depend on the source direction and the earth
rotation (the geometric delays), and for sufficiently narrow sub-bands
these delays become phase shifts.  Let $\bat_{q,k}$ denote this array
response vector towards the $q$th source at the $k$th snapshot.  We
assume that it is normalized such that $\bat_{q,k}^H\bat_{q,k}=1$.
In this notation, we use a tilde to denote parameters and related
matrices that depend on the `true' direction of the sources.  However,
in most of the paper we will work with parameters that are discretized
on a grid, in which case we will drop the tilde.  The grid points
correspond to the image pixels and do not necessary coincide with the
actual positions of the sources.

    Assuming an array that is otherwise calibrated, the received antenna
    signals $\by_k[n]$ can be modeled as
\begin{align}
\label{eq:samplemodel}
    \by_k[n]&=\bAtt_k\bxt[n]+\bn_k[n], \qquad n = 1, \cdots, N
\end{align}
    where $\bAtt_k$ is a $P \times Q$ matrix whose columns are the array
    response vectors $\bat_{q,k}$, $\bxt[n]$ is a $Q \times 1$ vector
    representing the signals from the sky, and $\bn_k[n]$ is a $P \times 1$
    vector modeling the noise.

    From the data, the system estimates covariance matrices (also known as 
    visibilities) of the input vector at each snapshot $k=1,\cdots,K$, as
\beq
    \bRh_{k} = \frac{1}{N} \sum_{n=1}^{N} \by_k[n]\by_k[n]^H, 
    \qquad k = 1, \cdots, K \,.
\eeq
    Since the received signals and noise are Gaussian, these covariance
    matrix estimates form sufficient statistics for the imaging problem
    \cite{leshem2000a}.  The covariance matrices are given by
\begin{equation}
    \bR_k=\EExp\{\by_k \by_k^H\}
\end{equation}
    for which the model is
\begin{equation}
\label{eq:cov_def}
    \bR_k = \bAtt_k \bSigmat\bAtt_k^H+\bR_{\bn,k},
\end{equation}
    where $\bSigmat=\EExp\{\bxt\bxt^H\}$ and
    $\bR_{\bn,k}=\EExp\{\bn_k\bn_k^H\}$ are the source and noise
    covariance matrices, respectively.  We have assumed that sky sources are
    stationary, and if we also assume that they are independent, we can model
    $\bSigmat=\diag(\bsigmat)$ where
\begin{equation}
    \bsigmat=
    \begin{bmatrix}\tilde{\sigma}_1 &,\dots, &\tilde{\sigma}_Q\end{bmatrix}^T 
\end{equation}
    represents the power of the sources.
    Vectorizing both sides of \eqref{eq:cov_def} we obtain
\begin{equation}
    \br_k = (\bAtt_k^* \circ \bAtt_k)\bsigmat
	+ \br_{\bn,k}
\end{equation}
    where $\br_k=\vect(\bR_k)$ and
    $\br_{\bn,k}=\vect(\bR_{\bn,k})$.  After stacking the vectorized
    covariances for all of the snapshots we obtain
\begin{equation}
\label{eq:vectRtotal}
    \br = \bPsit\bsigmat + \br_{\bn}
\end{equation}
    where
\begin{align}
    \br = \begin{bmatrix}\br_{1}\\ \vdots\\\br_{K}\end{bmatrix}
\,,\quad
    \bPsit=\begin{bmatrix} \bAtt_{1}^* \circ \bAtt_{1}
    \\ \vdots \\ \bAtt_{K}^* \circ \bAtt_{K}\end{bmatrix}
\,,\quad 
    \br_\bn= \begin{bmatrix}\br_{\bn,1}\\ \vdots\\\br_{\bn,K}\end{bmatrix}
    \,.
\end{align}
    Similarly we vectorize and stack the sample covariance matrices as
\begin{equation}
\label{def:brh}
    \brh_{k}=\vect(\bRh_{k})
\,,\qquad
    \brh = \begin{bmatrix}\brh_{1}\\ \vdots\\\brh_{K}\end{bmatrix} 
\,.
\end{equation}

    Instead of (\ref{eq:vectRtotal}), we can use the independence between the 
    time samples and also write the aggregate data model as
\begin{equation}
    \bR=\begin{bmatrix}
    \bR_1 &\dots &\zeros \\
    \vdots &\ddots &\zeros \\
    \zeros &\dots & \bR_K
    \end{bmatrix}
    = \sum_{q=1}^Q \tilde{\sigma}_q(\bI_K \circ \bAtt^q)
    (\bI_K \circ \bAtt^q)^H + \bR_{\bn} 
    \,,
\label{eq:blockdiagtilde}
\end{equation}
    where
\begin{equation}
    \bAtt^q=\begin{bmatrix} \bat_{q,1} & \dots & \bat_{q,K} \end{bmatrix}
    , \quad q=1,\cdots,Q
\end{equation}
    and
\begin{equation}
    \bR_\bn=\begin{bmatrix}
    \bR_{\bn,1} &\dots &\zeros \\
    \vdots &\ddots &\zeros \\
    \zeros &\dots & \bR_{\bn,K}
    \end{bmatrix}.
\end{equation}

\section{The Imaging Problem}
\label{sec:imagingp}

    Using the data model (\ref{eq:vectRtotal}), the imaging problem is
    to find the spatial power distribution $\bsigmat$ of the
    sources, along with their directions represented by the matrices
    $\bAtt_k$, from given sample covariance matrices $\bRh_k, \,
    k=1,\cdots,K$.  As the source locations are generally unknown, this is
    a complicated (non-linear) direction-of-arrival estimation problem. 

    The usual approach in radio astronomy is to define a grid for the image, and to assume
    that each pixel (grid location) contains a source.  In this case the
    source locations are known, and estimating the source powers is a
    linear problem, but for high-resolution images the number of sources
    may be very large.  The resulting linear estimation problem is
    often ill-conditioned unless additional constraints are posed.

\subsection{Gridded Imaging Model}
    After defining a grid for the image and assuming that a source
    exists for each pixel location, let $I$ denote the total number of
    sources (pixels), $\bsigma$ an $I\times 1$ vector containing the
    source powers, and $\bA_k$ ($k= 1, \cdots, K$) the $P\times I$ array
    response matrices for these sources.  Note that the $\bA_k$ are
    known, and that $\bsigma$ can be interpreted as a vectorized version
    of the image to be computed; we dropped the `tilde' in the notation to
    indicate the difference between the gridded pixel locations and the
    true (and unknown) source locations. 
    The $i$th column of $\bA_k$ is $\ba_{i,k}$, and
    similar to $\bat_{q,k}$ we assume that $\ba_{i,k}^H\ba_{i,k}=1$ for
    $i=1,\cdots,I$ and $k=1,\cdots,K$.

    The corresponding data model is now
\begin{equation}
\label{eq:cov_def_model}
    \bR_k=\bA_k \diag(\bsigma)\bA_k^H + \bR_{\bn,k} \,,
\end{equation}
    or in vectorized and stacked form (replacing (\ref{eq:vectRtotal}))
\begin{equation}
\label{eq:vectRImg}
    \br=\bAt\bsigma + \br_{\bn},
\end{equation}
    or in blockdiagonal form (replacing (\ref{eq:blockdiagtilde}))
\begin{equation}
\label{eq:blockdiag}
    \bR=\sum_{i=1}^I \sigma_i (\bI_K \circ \bA^i)
       (\bI_K \circ \bA^i)^H + \bR_{\bn}
       \,,
\end{equation}
    where 
\begin{equation}
    \bA_k=\begin{bmatrix} \ba_{1,k} , \cdots , \ba_{I,k} \end{bmatrix}
    ,\quad  k=1,\cdots, K
\end{equation}
\begin{equation}
\label{eq:bAt}
    \bAt=\begin{bmatrix} \left(\bA_{1}^* \circ \bA_{1} \right)^T,
    \cdots,\left( \bA_{K}^* \circ \bA_{K}\right)^T\end{bmatrix}^T
\end{equation}
\begin{equation}
    \bA^i=\begin{bmatrix} \ba_{i,1} & \dots & \ba_{i,K} \end{bmatrix},
    \quad  i=1,\cdots, I
    \,.
\end{equation}
    For a given observation $\brh$ in (\ref{def:brh}), image
    formation amounts to the estimation of $\bsigma$.  
    For a sufficiently fine grid, $\bsigma$ approximates the solution of the
    discrete source model.  However, as we will discuss later, working in
    the image domain leads to a gridding related noise floor.  This is
    solved by fine adaptation of the location of the sources and estimating
    the true locations in the visibility domain.

%
%

\subsection{Unconstrained Least Squares Images}
\label{subsec:UnWLS}
    If we ignore the term $\br_{\bn}$, then (\ref{eq:blockdiag})
    directly leads to Least Squares
    (LS) and Weighted Least Squares (WLS) estimates of $\bsigma$ 
    \cite{wijnholds2008}.  In particular, solving the imaging problem 
    with LS leads to the minimization problem
\begin{equation}
\label{eq:UCLS}
    \min_{{\bsigma}} \; \frac{1}{2K}\| \brh - \bAt\bsigma\|^2 \,.
\end{equation}
    It is straightforward to show that the solution to this problem is given
    by any $\bsigma$ that satisfies
\begin{equation}
\label{eq:imagingeq}
    \HLS\bsigma=\SMFDI
\end{equation}
    where we define the ``matched filter'' (MF, also known as the classical
    ``direct Fourier transform  dirty image'') as
\begin{equation}
\label{eq:mfdimg}
\SMFDI=\frac{1}{K}\bAt^H\brh=\frac{1}{K}\sum_k \vectdiag(\bA_k^H\bRh_k\bA_k),
\end{equation}
    and the deconvolution matrix $\HLS$ as
\begin{equation}
\label{eq:convMat}
\HLS=\frac{1}{K}\bAt^H\bAt=\frac{1}{K}\sum_k (\bA^T_k\bA^*_k)\odot (\bA^H_k\bA_k).
\end{equation}
    Similarly we can define the WLS minimization as
\begin{equation}
\label{eq:UWCLS}
    \min_{{\bsigma}} \frac{1}{2K}
   \| (\bRh^{-T/2}\otimes \bRh^{-1/2})(\brh - \bAt\bsigma) \|^2 \,,
\end{equation}
    where the weighting assumes Gaussian distributed observations.  The
    weighting improves the statistical properties of the estimates, and
    $\bRh$ is used instead of $\bR$ because it is available and gives
    asymptotically the same optimal results, i.e., convergence to maximum
    likelihood estimates \cite{Ottersten1998185}.
    The solution to this optimization is similar to the solution to the LS
    problem and is given by any $\bsigma$ that satisfies
\begin{equation}
\label{eq:imagingeqwls0}
    \HWLS\bsigma=\SWLSDI \,,
\end{equation}
    where
\begin{equation}
\label{eq:imagingeqwls}
    \SWLSDI=\frac{1}{K}\bAt^H(\bRh^{-T}\otimes \bRh^{-1})\brh
\end{equation}
    is the ``WLS dirty image'' and
\begin{equation}
\label{eq:imagingeqwls2}
    \HWLS=\frac{1}{K}\bAt^H(\bRh^{-T}\otimes \bRh^{-1})\bAt \,.
\end{equation}
   is the associated deconvolution operator.

\label{sec:bf}

    A connection to beamforming is obtained as follows.
    The $i$th pixel of the ``Matched Filter'' dirty image in equation 
    (\ref{eq:mfdimg}) can be written as
\[
    \SMFDIind{i} = \frac{1}{K} \sum_k \ba_{i,k}^H \bRh_k \ba_{i,k}
\]
    and if we replace $\ba_{i,k}/\sqrt{K}$ 
    by a more general ``beamformer'' $\bw_{i,k}$, 
    this can be generalized to a more general dirty image
\[
    \WTDIind{i} = \sum_k \bw_{i,k}^H \bRh_k \bw_{i,k}
\]
    Here, $\bw_{i,k}$ is called a beamformer because we can consider
    that it acts on the antenna vectors $\by_k[n]$ as $z_{i,k} =
    \bw_{i,k}^H \by_k[n]$, where $z_{i,k}$ is the output of the
    (direction-dependent) beamformer, and $\WTDIind{i} = 
    \sum_k \EExp\{|z_{i,k}|^2\}$ is interpreted as the total output power of the
    beamformer, summed over all snapshots.
    We will encounter several such beamformers in the rest of the paper.

\subsection{Preconditioned Weighted Least Squares}
\label{subsec:UnPWLS}
    If $\bAt$ has full column rank then $\HLS$ and $\HWLS$ are non-singular
    and there exists a unique solution to LS and WLS.  For example the
    solution to \eqref{eq:imagingeq} becomes
\beq
    \bsigma = \HLS^{-1}\SMFDI \,.
\eeq
    Unfortunately, if the number of pixels is large then $\HLS$ and $\HWLS$
    become ill-conditioned or even singular, so that
    \eqref{eq:imagingeq} and \eqref{eq:imagingeqwls0} have an infinite number
    of solutions \cite{wijnholds2008}.  Generally, we need to improve
    the conditioning of the deconvolution matrices and to find
    appropriate regularizations.

    One way to improve the conditioning of a matrix is by applying a
    preconditioner.  The most widely used and simplest preconditioner is the
    Jacobi preconditioner \cite{barrett1994} which, for any matrix $\bM$,
    is given by $[\diag(\bM)]^{-1}$.  Let $\bD_{\text{WLS}} =
    \diag(\HWLS)$, then by applying this
    preconditioner to $\HWLS$ we obtain
\begin{equation}
\label{eq:imagingeqwlspre}
    [\bD_{\text{WLS}}^{-1}\HWLS]\bsigma=\bD_{\text{WLS}}^{-1}\SWLSDI \,.
\end{equation}
	We take a closer look at $\bD_{\text{WLS}}^{-1}\SWLSDI$ for the case where $K=1$. In this case 
	\begin{align*}
	\HWLS&=(\bA_1^*\circ \bA_1)^H(\bRh_1^{-T} \otimes \bRh^{-1}_1)(\bA_1^*\circ \bA_1)\\
		&=(\bA^T\bRh_1^{-T}\bA_1^*)\odot(\bA_1^H\bRh_1^{-1}\bA_1)
	\end{align*}
	and
	\[
		\bD_{\text{WLS}}^{-1}=\begin{bmatrix}
		\frac{1}{(\ba_{1,1}^H\bRh_1^{-1}\ba_{1,1})^2} & & \\
		& \ddots & \\
		& & \frac{1}{(\ba_{I,1}^H\bRh_1^{-1}\ba_{I,1})^2}
		\end{bmatrix}.
	\]
	This means that
	\begin{align*}
		\bD_{\text{WLS}}^{-1}\SWLSDI&=\bD_{\text{WLS}}^{-1}(\bRh_1^{-T} \otimes \bRh^{-1}_1)(\bA_1^*\circ \bA_1)^H\brh_1 \\
		&=(\bRh_1^{-T}\bA_1^*\bD_{\text{WLS}}^{-1/2} \circ \bRh_1^{-1}\bA_1\bD_{\text{WLS}}^{-1/2})^H\brh_1
	\end{align*}
    which is equivalent to a dirty image that is obtained by applying a beamformer of the form
\begin{equation}
    \bw_i = \frac{1}{\ba_{i,1}^H\bRh_1^{-1}\ba_{i,1}}\bRh_1^{-1}\ba_{i,1}
\end{equation}
    to both sides of $\bRh_{1}$ and stacking the results,
    $\hat{\sigma}_i = \bw_i^H\bRh_{1}\bw_i$, of each pixel into a
    vector.  This beamformer is known in array processing as 
    the Minimum Variance Distortionless Response (MVDR) beamformer
    \cite{capon69}, and the corresponding MVDR dirty image was
    introduced in the radio astronomy context in \cite{leshem2000a}.


\subsection{Bounds on the Image}
\label{power_constraint}

    Another approach to improve the conditioning of a problem is to
    introduce appropriate constraints on the solution.
    Typically, image formation algorithms exploit external information
    regarding the image in order to regularize the ill-posed problem.  For
    example maximum entropy techniques \cite{frieden72, gull78} impose a
    smoothness condition on the image while the CLEAN algorithm
    \cite{hogbom74} exploits a point source model wherein most of the image
    is empty, and this has recently been connected to sparse optimization
    techniques \cite{wiaux09}.

    A lower bound on the image is almost trivial: each pixel in the image
    represents the power coming from a certain direction, hence is
    non-negative.  This leads to a lower bound $\bsigma\geq \zeros$.  Such
    a non-negativity constraint has been studied for example in
    \cite{briggs95}, resulting in a non-negative LS (NNLS) problem 
\beq
\begin{array}{l}
\label{eq:NNLS}
    \displaystyle\min_{{\bsigma}}\, \frac{1}{2K}\| \brh - \bAt\bsigma\|^2 \\
    \hbox{subject to } \vzero \le \bsigma
\ena
\eeq
    A second constraint follows if we also know an upper bound $\bgamma$
    such that $\bsigma \leq \bgamma$, which will bound the pixel powers
    from above.  We will propose several choices for $\bgamma$.

    Actual dirty images are based on the sample covariance matrix $\bRh$
    and hence they are random variables.
    By closer inspection of the $i$th pixel of the MF dirty image $\SMFDI$,
    we note that its expected value is given by
\[
    \MFDIind{i} =\frac{1}{K}\sum_k\ba^H_{i,k} \bR_k \ba_{i,k} \,.
\]
     Using 
     \begin{equation}
\label{eq:longa}
    \ba_i = \vect(\bA^i) = 
    \begin{bmatrix}\ba_{i,1}^T & \dots &  \ba_{i,K}^T \end{bmatrix}^T,
\end{equation}
and the normalization $\ba_{i,k}^H \ba_{i,k} = 1$, we obtain
\beq
\label{eq:MFupper}
    \MFDIind{i} =\frac{1}{K}\ba_{i}^H\bR\ba_i=\sigma_{i}+\frac{1}{K}\ba_{i}^H\bR_r\ba_i,
\eeq
    where (cf.\ (\ref{eq:blockdiag}))
\begin{equation}
    \bR_{r}=\sum_{j \neq i} \sigma_j 
    (\bI_K \circ \bA^j)(\bI_K \circ \bA^j)^H + \bR_{\bn}
\end{equation}
    is the contribution of all other sources and the noise. Note that
    $\bR_r$ is positive-(semi)definite.  Thus, (\ref{eq:MFupper}) implies
    $\MFDIind{i} \ge \sigma_{i}$ which means that the expected value of the
    MF dirty image forms an upper bound for the desired image, or
\begin{equation}
\label{eq:mfupper}
    \bsigma \leq \MFDI \,.
\end{equation}

    Now let us define a beamformer
    $\bw_{\text{MF},i}=\frac{1}{\sqrt{K}}\ba_i$, then we observe that each
    pixel in the MF dirty image is the output of this beamformer:
\begin{equation}
\MFDIind{i}=\bw_{\text{MF},i}^H\bR\bw_{\text{MF},i}.
\end{equation}
    As indicated in Sec.\ \ref{sec:bf}, we can extend this concept to a
    more general beamformer $\bw_{i}$.  The output power of
    this beamformer, in the direction of the $i$th pixel, becomes
\beq
\label{eq:beamformer}
    \WTDIind{i}= \bw_{i}^H\bR\bw_{i}=\sigma_{i}\bw_i^H(\bI_K \circ
    \bA^i)(\bI_K \circ \bA^i)^H\bw_i+ \bw_{i}^H\bR_r\bw_{i} \,.
\eeq
    If we require that 
\beq
\label{eq:mvdrconst_mulit}
    \bw_i^H(\bI_K \circ \bA^i)(\bI_K \circ \bA^i)^H\bw_i=1
\eeq
    we have
\beq
\label{eq:MVDRmin}
    \WTDIind{i}=\sigma_{i}+\bw_{i}^H\bR_{r}\bw_{i} \,.
\eeq
    As before, the fact that $\bR_r$ is positive definite implies that 
\beq
\label{eq:wupper}
    \sigma_{i} \le \WTDIind{i} \,.
\eeq
    We can easily verify that $\bw_{\text{MF},i}$ satisfies
    \eqref{eq:mvdrconst_mulit} and hence $\MFDIind{i}$ is a specific
    upper bound.  A question which arises at this point is: What is the
    tightest upper bound for $\sigma_i$ that we can construct using
    linear beamforming?  We can translate this to the following
    optimization question:
\begin{align}
    \sigma_{\text{opt},i}&=\min_{\bw_i} \bw_{i}^H\bR\bw_{i} 
    \\ & 
    \text{s.t.~}  \bw_i^H(\bI_K \circ \bA^i)(\bI_K \circ \bA^i)^H\bw_i=1
\notag
\end{align}
    where $\sigma_{\text{opt},i}$ would be this tightest upper bound.
    This problem can be solved (Appendix. \ref{appen:assc}): the tightest upper bound is given by 
\begin{equation}
\label{eq:ASSC0}
    \sigma_{\text{opt},i}=\min_k
    \left(\frac{1}{\ba_{i,k}^H\bR_k^{-1}\ba_{i,k}}\right),
\end{equation}
    and the beamformer that achieves this was called the adaptive
    selective sidelobe canceller (ASSC) in \cite{levanda2013}.  One
    problem with using this result in practice is that
    $\sigma_{\text{opt},i}$ depends on a single snapshot.  This means
    that there is a variance-bias trade-off when we have a sample
    covariance matrix $\bRh$ instead of the true covariance matrix
    $\bR$.  An analysis of this problem and various solutions for it are
    discussed in \cite{levanda2013}.

    To reduce the variance we will tolerate an increase of the
    bound with respect to the tightest, however we would like our result to
    be tighter than the MF dirty image.  For this reason we suggest to find a
    beamformer that instead of (\ref{eq:mvdrconst_mulit}) satisfies
    the slightly different normalization constraint
\begin{equation}
    \bw_i^H\ba_{i}=\sqrt{K}  \,.
\end{equation}
    We will show that the expected value of the resulting dirty image 
    constitutes a larger upper bound than the ASSC
    \eqref{eq:ASSC0}, but because the output power of this beamformer
    depends on more than one snapshot it will have a lower variance than
    ASSC, so that it is more robust in practice.

    With this constraint, the beamforming problem is 
\begin{align}
\label{eq:mvdrproblem0}
    \bw_i&=\arg\min_{\bw_i} \bw_i^H\bR\bw_i \\
	 & \text{s.t.~}  \bw_i^H\ba_i=\sqrt{K}
    \notag
\end{align}
    which is recognized as the classical 
    minimum variance distortionless response
    (MVDR) beamforming problem \cite{capon69}. Thus, the solution 
    is given in closed form as 
\begin{equation}
\label{eq:wmvdr}
    \bw_{\text{MVDR},i} = 
    \frac{\sqrt{K}}{\ba_{i}^H\bR^{-1}\ba_{i}} \bR^{-1}\ba_{i}
\end{equation}
    and the resulting MVDR dirty image is
\begin{align}
    \MVDRDIind{i}
    & = \bw_{\text{MVDR},i}^H \bR \bw_{\text{MVDR},i}
    \notag \\
    & = \frac{K\sum_k\ba_{i,k}^H\bR_{k}^{-1}\ba_{i,k}}
    {\left(\sum_k\ba_{i,k}^H\bR_{k}^{-1}\ba_{i,k}\right)^2} 
    \notag \\
    & = \frac{1}{\frac{1}{K}\sum_k\ba_{i,k}^H\bR_{k}^{-1}\ba_{i,k}} \,.
\end{align}
    Interestingly, for $K=1$ this is the same image as we obtained earlier by applying a Jacobi preconditioner to the WLS problem.
    To demonstrate that this image is still an upper bound we show that
\beq
    \alpha := \bw_i^H(\bI_K \circ \bA^i)(\bI_K \circ \bA^i)^H\bw_i \geq 1
    \,.
\eeq
    Indeed, inserting \eqref{eq:wmvdr} into this inequality gives
    \begin{equation}
\begin{array}{l}
\label{eq:MVDRUB2}
    K\frac{\ba_i^H\bR^{-1}(\bI_K \circ \bA^i)(\bI_K \circ \bA^i)^H\bR^{-1}\ba_i}{(\ba_i^H\bR^{-1}\ba_i)^2}  \\ = K \frac{\sum_k (\ba_{i,k}^H\bR_k^{-1}\ba_{i,k})^2}{\left(\sum_k \ba_{i,k}^H\bR_k^{-1}\ba_{i,k}\right)^2}\\
    = K\frac{\bh^T\bh}{\bh^T\ones_K\ones_K^T\bh} \geq K\frac{1}{\lambda_{\text{max}}(\ones_K\ones_K^T)}=1,
\end{array}
\end{equation}
    where $\bh=(\bI_K \circ \bA^i)^H\bR^{-1}\ba_i$ is a $K \times 1$
    vector with entries $h_k=\ba_{i,k}^H\bR_k^{-1}\ba_{i,k}$ and
    $\lambda_{\text{max}}(\cdot)$ is the largest eigenvalue of of the
    argument matrix.  Hence, a similar reasoning as in (\ref{eq:beamformer}) 
    gives
\[
    \MVDRDIind{i} = \alpha \sigma_i + \bw_{\text{MVDR},i}^H \bR_r
    \bw_{\text{MVDR},i} \ge \sigma_i
\]
    which is
\beq
\label{eq:mvdr_upper}
    \bsigma \le \MVDRDI \,.
\eeq
    Note that $\bw_{\text{MF},i}$ also satisfies the constraint in
    (\ref{eq:mvdrproblem0}), i.e.\ $\bw_{\text{MF},i}^H \ba_i=
    \sqrt{K}$, but does not necessary minimize the output power $\bw_i^H
    \bR \bw_i$, therefore the MVDR dirty image is smaller than the MF
    dirty image: $\MVDRDI \le \MFDI$.  Thus it is a tighter upper bound.
    This relation also holds if $\bR$ is replaced by the sample
    covariance $\bRh$.

\subsection*{Estimation of the Upper Bound from Noisy Data}

    The upper bounds \eqref{eq:mfupper} and (\ref{eq:mvdr_upper}) assume
    that we know the true covariance matrix $\bR$.  However in practice we
    only measure $\bRh$ which is subject to statistical fluctuations.
    Choosing a confidence level of $6$ times the standard deviation of the
    dirty images ensures that the upper bound will hold with probability
    99.9\%.
    This leads to an increase of the upper bound by a factor $1+\alpha$ where $\alpha >
    0$ is chosen such that
\begin{equation}
\bsigma \leq (1+\alpha) ~ \SMFDI.
\end{equation}
    Similarly, for the MVDR dirty image the constraint based on $\bRh$ is
\begin{equation}
\bsigma \leq (1+\alpha) ~ \UMVDRDI
\end{equation}
where
\begin{equation}
\UMVDRDIind{i} =  \frac{C}{\frac{1}{K}\sum_k\ba_{i,k}\bRh_{k}^{-1}\ba_{i,k}}
\end{equation}
    is an unbiased estimate of the MVDR dirty image, and
\begin{equation}
C=\frac{N}{N-p}
\end{equation}
    is a bias correction constant.  With some algebra the unbiased estimate
    can be written in vector form as
\begin{equation}
\label{eq:UBMVDRV}
    \UMVDRDI=\bD^{-1}\bAt^H(\bRh^{-T} \otimes \bRh^{-1})\brh,
\end{equation}
    where
\begin{equation}
\label{eq:D}
\bD =\frac{1}{KC}\diag^2\left( \bA^H\bRh^{-1}\bA\right),
\end{equation}
    and
\begin{align}
\bA&=\begin{bmatrix}\bA_1^T & \dots &\bA_K^T\end{bmatrix}^T \notag \\
&=\begin{bmatrix} \ba_1 & \dots & \ba_I \end{bmatrix}.
\end{align}
    The exact choice of $\alpha$ and $C$ are discussed in Appendix \ref{appedix:1}.


\subsection{Constrained Least Squares Imaging}
\label{sec:LS}

    Now that we have lower and upper bounds on the image, we can use these
    as constraints in the LS imaging problem to provide a regularization.
    The resulting constrained LS (CLS) imaging problem is
\beq
\label{eq:CLS}
\begin{array}{l}
	 \displaystyle\min_{{\bsigma}} \frac{1}{2K}\| \brh - \bAt\bsigma\|^2  \\
	\hbox{s.t. } \vzero \le \bsigma \le \bgamma 
    \end{array}
\eeq
    where $\bgamma$ can be chosen either as $\bgamma=\SMFDI$ for the MF dirty
    image or $\bgamma=\UMVDRDI$ for the MVDR dirty image.

    The improvements to the unconstrained LS problem that where discussed in
    Sec.\ \ref{subsec:UnWLS} are still applicable.  The extension to WLS
    leads to the cost function

\beq
\label{def:vecfWLS}
    f_{\text{WLS}}(\bsigma)=\frac{1}{2}\| (\bRh^{-T/2} \otimes \bRh^{-1/2})
    \left(\brh-\bAt\bsigma\right)\|^2
    \,.
\eeq
    The constrained WLS problem is then given by
\beq
\label{eq:WLS}
    \begin{array}{ll}
    \displaystyle\min_{{\bsigma}} f_{\text{WLS}}(\bsigma)  \\
    \hbox{s.t. } \vzero \le \bsigma \le \bgamma \,.
    \end{array}
\eeq
	We also recommend to include a preconditioner which, as was shown in Sec.\ref{subsec:UnPWLS}, relates the WLS to the MVDR dirty image. However, because of the inequality constraits, \eqref{eq:WLS} does not have a closed form solution and it is solved by an iterative algorithm. In order to have the relation between WLS and MVDR dirty image during the iterations we introduce a change of variable of the form $\PWLSsigma=\bD \bsigma$, where $\PWLSsigma$ is the new variable for the preconditioned problem and the diagonal matrix $\bD$ is given in \eqref{eq:D}. The
    resulting constrained preconditioned WLS (PWLS) optimization problem is 
\beq
\begin{array}{rl}
\label{eq:preWLS}
    \PWLSsigma=& \displaystyle\arg \min_{\PWLSsigma} 
	\frac{1}{2}\| (\bRh^{-T/2} \otimes \bRh^{-1/2})
	\left(\brh-\bAt\bD^{-1}\PWLSsigma\right)\|^2  \\
    & \hbox{s.t. } \vzero \le\PWLSsigma\le \bD \bgamma 

\end{array}
\eeq
    and the final image is found by setting $\bsigma=\bD^{-1} \PWLSsigma$.
    (Here we used that $\bD$ is a positive diagonal matrix so that
    the transformation to an upper bound for $\PWLSsigma$ is correct.)  
    Interestingly, the dirty image that follows from the (unconstrained)
    Weighted Least Squares part
    of the problem is given by the MVDR image $\UMVDRDI$ in (\ref{eq:UBMVDRV}).

\section{Constrained optimization using an active set method}
\label{sec:COPAS}

    The constrained imaging formulated in the previous section requires
    the numerical solution of the optimization problems (\ref{eq:CLS}) or
    (\ref{eq:preWLS}).  The problem is classified as a positive definite
    quadratic program with simple bounds, this is a special case of a
    convex optimization problem with linear inequality constraints, and we
    can follow standard approaches to find a solution
    \cite{Gill81,boyd2004}.

    For an unconstrained optimization problem, the gradient of the cost
    function calculated at the solution must vanish.  If in an iterative
    process we are not yet at the optimum, the gradient is used to update the
    current solution.
    For constrained optimization, the constraints are usually added to the
    cost function using (unknown) Lagrange multipliers that need to be
    estimated along with the solution.  At the solution, part of the
    gradient of the cost function is not zero but related to the nonzero
    Lagrange multipliers. For inequality constraints, the sign of the Lagrange
    multipliers plays an important role.
    
    In this Section, we use an approach called the active
    set method to solve the constrained optimization problem.

\subsection{Characterization of the Optimum}

    Let $\bar{\bsigma}$ be the solution to the optimization problem
    (\ref{eq:CLS}) or (\ref{eq:preWLS}).   An image is called feasible if
    it satisfies the bounds $\bsigma \ge \zeros$ and $-\bsigma \ge -\bgamma$.
    At the optimum, some pixels may satisfy a bound with equality, and
    these are called the ``active'' pixels. 

    We will use the following notation. 
    For any feasible image $\bsigma$, let
\begin{align}
    \MCL(\bsigma)&=\{i\,|\,\sigma_{i} = 0\}\\
    \MCU(\bsigma)&=\{i\,|\,\sigma_{i} = \gamma_i\}\\
    \MCA(\bsigma)&=\MCL(\bsigma) \cup \MCU(\bsigma)\\
    \MCF(\bsigma)&=\MCI \setminus \MCA(\bsigma)\,.
\end{align}
    $\MCI = \{1, \cdots, I\}$ is the set of all pixel indices,
    $\MCL(\bsigma)$ is the set where the lower bound is active, i.e.,
    the pixel value is $0$.  $\MCU(\bsigma)$ is the set of pixels which
    attain the upper bound.  $\MCA(\bsigma)$ is the set of all pixels
    where one of the constraints is active, these are the
    active pixels.  Finally, the free set $\MCF(\bsigma)$ is the set of
    pixels $i$ which have values strictly between $0$ and $\gamma_i$.
    Further, for any vector $\bv = [v_i]$, let $\bv_\MCF$ correspond to the
    subvector with indices $i \in \MCF$, and similarly define $\bv_\MCL$
    and $\bv_\MCU$.  We will write $\bv = \bv_\MCF \oplus \bv_\MCL \oplus
    \bv_\MCU$.

    Let $\bar{\bsigma}$ be the optimum, and let $\bar{\bg} =
    \bg(\bar{\bsigma})$ be the gradient of the cost function at this point.
    Define the free sets and active sets $\MCF, \MCL, \MCU$ at $\bar{\bsigma}$.
    We can write $\bar{\bg} = \bar{\bg}_\MCF \oplus \bar{\bg}_\MCL \oplus
    \bar{\bg}_\MCU$.
    Associated with the active pixels of $\bar{\bsigma}$ is a vector
    $\bar{\blambda} = \bar{\blambda}_\MCL \oplus \bar{\blambda}_\MCU$ 
    of Lagrange multipliers. Optimization theory \cite{Gill81} tells us
    that the optimum $\bar{\bsigma}$
    is characterized by the following conditions:
\begin{align}
\label{eq:gradFcond}
    \bg_{\MCF}(\bar{\bsigma})&= \zeros \\
\label{eq:etaL}
    \bar{\blambda}_\MCL = \bar{\bg}_{\MCL} &\ge \zeros \\
\label{eq:etaU}
    \bar{\blambda}_\MCU = -\bar{\bg}_{\MCU}&\ge \zeros \,.
\end{align}
    Thus, the part of the gradient corresponding to the free set is zero, 
    but the part of the gradient corresponding to the 
    active pixels is not necessarily zero.  Since we have simple bounds,
    this part becomes equal to the Lagrange multipliers $\bar{\blambda}_\MCL$
    and $-\bar{\blambda}_\MCU$ (the negative sign is caused by the
    condition $-\bsigma_\MCU \ge -\bgamma_\MCU$).  
    The condition $\blambda \ge \zeros$ is crucial: a negative
    Lagrange multiplier would indicate that there exists a feasible direction
    of descent $\bp$ for which a small step into that direction,
    $\bar{\bsigma} + \mu \bp$, has a lower cost and still satisfies the
    constraints, thus contradicting optimality of $\bar{\bsigma}$
    \cite{Gill81}.

    ``Active set'' algorithms consider that if the true active set at the
    solution would be known, the optimization problem with inequality
    constraints reduces to an optimization with equality constraints,
\begin{align}
\label{eq:subproblem}
    \bz = & \arg\min_{\bsigma}f(\bsigma)\\
    \text{s.t.} ~& \bsigma_{{\MCL}}=\zeros\,,\; 
    \bsigma_{{\MCU}}=\bgamma_{\MCU} \,. \notag
\end{align}
    Since we can substitute the values of the active pixels into $\bsigma$, 
    the problem becomes a standard unconstrained LS problem with a reduced
    dimension: only $\bar{\bsigma}_\MCF$ needs to be estimated. 
    Specifically, for CLS the unconstrained subproblem is formulated as
\begin{equation}
\label{eq:subproblemLS}
    f(\bsigma) = 
    \frac{1}{2K}\|\bb_{\text{LS}}-\bAt_{\MCF}\bsigma_{\MCF} \|^2
\end{equation}
    where
\begin{equation}
\label{eq:bsubproblemLS}
    \bb_{\text{LS}}=\brh-\bAt_{\MCU}\bsigma_{\MCU}.
\end{equation}
    Similarly for PWLS we have
\begin{equation}
\label{eq:subproblemMVDR}
    f(\PWLSsigma) = 
    \frac{1}{2}\left \|\bb_{\text{PWLS}}-
    \left (\bRh^{-T/2}\otimes \bRh^{-1/2} \right)
    (\bAt\bD^{-1})_{\MCF}\PWLSsigma_{\MCF}\right \|^2
\end{equation}
    where
\begin{equation}
\label{eq:bsubproblemMVDR}
    \bb_{\text{PWLS}}=\left (\bRh^{-T/2}\otimes \bRh^{-1/2} \right)
    (\brh-(\bAt\bD^{-1})_{\MCU}\PWLSsigma_{\MCU})
\end{equation}
    In both cases, closed form solutions can be found, and we will discuss
    a suitable Krylov-based algorithm for this in Sec.\ \ref{sec:krylov}.

    Hence the essence of the constrained optimization problem is to find $\MCL$,
    $\MCU$ and $\MCF$.  In the literature algorithms for this are called
    \emph{active set methods}, and we propose a suitable algorithm in Sec.\
    \ref{sec:active}.

\subsection{Gradients}

    We first derive expressions for the gradients required for each of
    the unconstrained subproblems (\ref{eq:subproblemLS}) and
    (\ref{eq:subproblemMVDR}).
    Generically, a WLS cost function (as function of a 
    real-valued parameter vector $\btheta$) has the form
\begin{equation}
    f(\btheta)_{\text{WLS}} 
    = \beta \|\bG^{1/2}\bc(\btheta)\|^2
    = \beta\bc(\btheta)^H\bG\bc(\btheta)
\end{equation}
    where $\bG$ is a Hermitian weighting matrix and $\beta$ is a scalar.
    The gradient of this function is
\begin{equation}
\label{eq:gWLS}
    \bg(\btheta) = 
    2\beta\left(\frac{\partial \bc}{\partial \btheta^T}\right)^H\bG\bc
    \,.
\end{equation}
    For LS we have $\btheta=\bsigma$, $\bc=\brh-\bAt\bsigma$,
    $\beta=\frac{1}{2K}$ and $\bG=\bI$.  This leads to
\begin{align}
\label{eq:gls}
    \bg_{\text{LS}}(\bsigma) & =-\frac{1}{K}\bAt^H(\brh-\bAt\bsigma) \notag \\
    &=\bH_{\text{LS}}\bsigma-\SMFDI.
\end{align}
    For PWLS, $\btheta=\PWLSsigma$,
    $\bc=\brh-\bAt\bD^{-1}\PWLSsigma$, $\beta=\frac{1}{2}$ and
    $\bG=\bRh^{-T} \otimes \bRh^{-1}$.  Substituting into \eqref{eq:gWLS} we
    obtain
\begin{align}
\label{eq:gmvdr}
    \bg_{\text{PWLS}}(\PWLSsigma)
    &=-\bD^{-1}\bAt^H(\bRh^{-T} \otimes \bRh^{-1})(\brh-\bAt\bD^{-1}\PWLSsigma)
    \notag \\
    &=\bH_{\text{PWLS}}\PWLSsigma-\UMVDRDI
\end{align}
    where
\begin{equation}
\bH_{\text{PWLS}} = \bD^{-1}\bAt^H(\bRh^{-T} \otimes \bRh^{-1})\bAt\bD^{-1},
\end{equation}
    and we used \eqref{eq:UBMVDRV}. 

    An interesting observation is that the gradients can be interpreted
    as residual images obtained by subtracting the dirty image from a
    convolved model image.  This will at a later point allow us to
    relate the active set method to sequential source removing
    techniques.

%
%

\subsection{Active Set Methods}
\label{sec:active}

    In this section, we describe the steps needed to find the sets
    $\MCL$, $\MCU$ and $\MCF$, and the solution.  We follow the template
    algorithm proposed in \cite{Gill81}.  The algorithm is an iterative
    technique where we gradually improve on an image.  Let the image
    at iteration $j$ be denoted by $\bsigma^{(j)}$ where $j=1,2,\cdots$,
    and we always ensure this is a feasible solution (satisfies $\zeros
    \le \bsigma^{(j)} \le \bgamma$).  The corresponding gradient is the
    vector $\bg = \bg(\bsigma^{(j)})$, and the current estimate of the
    Lagrange multipliers $\blambda$ is obtained from $\bg$ using
    (\ref{eq:etaL}), (\ref{eq:etaU}).  The sets $\MCL$, $\MCU$ and
    $\MCF$ are current estimates that are not yet necessarily equal to
    the true sets.

    If this image is not yet the true solution, it means that one of the
    conditions in (\ref{eq:gradFcond})--(\ref{eq:etaU}) is violated. If the
    gradient corresponding to the free set is not yet zero ($\bg_\MCF \neq
    \zeros$), then this is remedied by recomputing
    the image from the essentially unconstrained subproblem
    (\ref{eq:subproblem}).  It may also happen that some entries of
    $\blambda$ are negative.  This implies that we do not yet have the
    correct sets $\MCL$, $\MCU$ and $\MCF$.  Suppose $\lambda_i < 0$.
    The connection of $\lambda_i$ to the gradient indicates that the
    cost function can be reduced in that dimension without violating any
    constraints \cite{Gill81}, at the same time making that pixel not
    active anymore.  Thus we remove the $i$th pixel from the active set,
    add it to the free set, and recompute the image with the new
    equality constraints using (\ref{eq:subproblem}).   As discussed later,
    a threshold $\epsilon$ is needed in the test for negativity of
    $\lambda_i$ and therefore this step is called the ``detection
    problem''.

    Table \ref{table:AS} summarizes the resulting active set algorithm and
    describes how the solution $\bz$ to the subproblem is used at each
    iteration.  Some efficiency is obtained by not computing the 
    complete gradient $\bg$ at every iteration, but
    only the parts corresponding to $\MCL, \MCU$, when they are needed.
    For the part corresponding $\MCF$, we use a flag that indicates whether
    $\bg_\MCF$ is zero or not.

\newcommand{\bgisnotzero}{{\em Freegradient-isnotzero}}
\begin{table}
\caption{Constrained LS Imaging Using Active Sets}
\label{table:AS}
\begin{algorithmic}[1]
\STATE Initialize: set the initial image $\bsigma^{(0)} = \zeros$, $j = 0$, 
     set the free set $\MCF = \emptyset$, and $\MCL, \MCU$ accordingly
\STATE Set the flag \bgisnotzero\ := True
\WHILE{\bgisnotzero\ {\bf or}\ $\lambda_{\min} < 0$}
    \IF {\bgisnotzero}
        \STATE Let $\bz$ be the solution of the unconstrained subproblem
	(\ref{eq:subproblem})
	\IF {$\bz$ is feasible}
	    \STATE Update the image: $\bsigma_\MCF^{(j+1)}= \bz$
	    \STATE Set \bgisnotzero\ := False
	    \STATE Compute the ``active'' part of the gradient and
		estimate the Lagrange multipliers
	    \STATE Let $\lambda_{\min}$ be the smallest Lagrange multiplier
		and $i_{\min}$ the corresponding pixel index
	\ELSE
	    \STATE Compute the direction of descent 
	        $\bp=\bz-\bsigma_{\MCF}^{(j)}$
	    \STATE Compute the maximum feasible nonnegative step-size
		$\mu_{\max}$ and let $i$ be the corresponding pixel index
		that will attain a bound
            \STATE Update the image: $\bsigma_{\MCF}^{(j+1)}=
	         \bsigma_{\MCF}^{(j)}+\mu_{\max}\bp$
	    \STATE Add a constraint: move $i$ from the free set $\MCF$ to $\MCL$
		or $\MCU$
	    \STATE Set \bgisnotzero\ := True
	\ENDIF
	\STATE Increase the image index: $j := j+1$
    \ELSE
	\STATE Delete a constraint: move $i_{\min}$ from $\MCL$ or $\MCU$ to 
	   the free set $\MCF$
	\STATE Set \bgisnotzero\ := True
    \ENDIF
\ENDWHILE
\end{algorithmic}
\end{table}

    In line 1, the iterative process is initialized.  This can be done
    in many ways.  As long as the initial image lies within the feasible
    region ($\zeros \leq \bsigma^{(0)} \leq \bgamma$), the
    algorithm will converge to a constrained solution.
    We can simply initialize by $\bsigma^{(0)} = \zeros$.

    Line 3 is a test for convergence, corresponding to the conditions
    (\ref{eq:gradFcond})--(\ref{eq:etaU}). The loop is followed while
    a condition is violated.

    If $\bg_\MCF$ is not zero, then in line 5 the unconstrained subproblem
    (\ref{eq:subproblem}) is solved. If this solution $\bz$ satisfies the
    feasibility constraints, then it is kept, the image is updated
    accordingly, and the gradient is estimated at the new solution (only
    $\lambda_{\min} = \min(\blambda)$ is needed, along with the corresponding
    pixel index).

    If $\bz$ is not feasible, then in line 12-16 
    we try to move into the direction of
    $\bz$ as far as possible. The 
    direction of descent is $\bp = \bz - \bsigma_\MCF^{(j)}$, 
    and the update will be 
    $\bsigma_{\MCF}^{(j+1)}= \bsigma_{\MCF}^{(j)}+\mu\bp$, where $\mu$ is a
    non-negative step size. The $i$th pixel will hit a bound if either 
    $\sigma_i^{(j)} + \mu p_i = 0$ or $\sigma_i^{(j)} + \mu p_i = \gamma_i$, i.e., if
\begin{equation}
\label{eq:mumax}
\mu_i=\max
   \left(-\frac{\sigma_i^{(j)}}{p_i} ,
        \frac{\gamma_i-\sigma_i^{(j)}}{p_i}\right)
\end{equation}
    (note that $\mu_i$ is non-negative). Then the maximal feasible step
    size towards a constraint is given by $\mu_{\max} =\min(\mu_i)$,
    for $i \in \MCF$.  The corresponding pixel index is removed from $\MCF$
    and added to $\MCL$ or $\MCU$.

    If in line 3 the gradient satisfied $\bg_\MCF = \zeros$ but a Lagrange
    multiplier was negative, we delete the corresponding constraint
    and add this pixel index to the free set (line 20).  After this, the
    loop is entered again with the new constraint sets.

    Suppose we initialize the algorithm with $\bsigma^{(0)}=\zeros$, then
    all pixel indices will be in
    the set $\MCL$, and the free set is empty.  
    During the first iteration $\bsigma_\MCF$ remains empty but the
    gradient is computed (line 9). Equations 
    \eqref{eq:gls} and \eqref{eq:gmvdr} show that it will be
    equal to the negated dirty image.
    Thus the minimum of the Lagrange multipliers $\lambda_{\min}$ 
    will be the current strongest source in the dirty image and
    it will be added to the free set when the loop is entered again.
    This shows that the method as described above will lead to a
    sequential source removal technique similar to CLEAN.  In
    particular, the PWLS cost function \eqref{eq:gmvdr} relates to LS-MVI
    \cite{bendavid08}, which applies CLEAN-like steps to the MVDR
    dirty image.

    In line 3, we try to detect if a pixel should be added to the free set
    ($\lambda_{\min} < 0$). Note that $\blambda$ follows from the gradient, 
    \eqref{eq:gls} or \eqref{eq:gmvdr}, which is a random variable.
    We should avoid the occurence of a ``false alarm'', because it will
    lead to overfitting the noise.
    Therefore, the test should be replaced by $\lambda_{\min} < -\epsilon$, 
    where $\epsilon> 0$ is a suitable detection threshold.
    Because the gradients are estimated using dirty images, they share
    the same statistics (the variance of the other component in \eqref{eq:gls} and \eqref{eq:gmvdr} is much
    smaller).  To reach a desired false alarm rate, we propose to choose
    $\epsilon$ proportional to the standard deviation of the $i$th pixel on the
    corresponding dirty image for the given cost function.  (How to estimate
    the standard deviation of the dirty images and the threshold is
    discussed in Appendix \ref{appedix:1}).  Choosing $\epsilon$ to be $6$ times the
    standard deviation ensures a false alarm of ${}< 0.1\%$ over the
    complete image.

    The use of this statistic improves the detection and hence the estimates
    greatly, however the correct detection also depends on the quality of the
    power estimates in the previous iterations.  If a strong source is
    off-grid, the source is usually underestimated, this leads to a biased
    estimation of the gradient and the Lagrange multipliers, which in turn
    leads to inclusion of pixels that are not real sources.  In the next
    section we describe one possible solution for this case.

%
%

\subsection{Strong Off-Grid Sources}
\label{sec:strong_sources}

    The mismatch between $\bAt$ and the unknown $\bPsit$ results in an
    underestimation of source powers, which means that the remaining power
    contribution of that source produces bias and possible artifacts in the
    image.  In order to achieve high dynamic ranges we suggest finding a
    grid correction for the pixels in the free set $\MCF$.

    So far we have not introduced a specific model for the elements in the
    matrix $\bAtt_{k}$, but in order to be able to correct for these
    gridding mismatches we assume that the array is at least calibrated for
    gains such that we can model the columns of this steering matrix as
\begin{equation}
\bat_{q,k}=\frac{1}{\sqrt{P}}e^{\frac{j 2\pi}{\lambda}\bXi^T\bQ_{k}(L,B)\tilde{\bbeta}_q}
\end{equation}
    where $\bXi$ is a $3 \times P$ matrix containing the position of each
    receiving element, $\bQ_k$ is a $3 \times 3$ rotation matrix that
    accounts for the earth rotations and depends on time and the observer's
    longitude $L$ and latitude $B$, $\tilde{\bbeta}_q$ is a $3 \times 1$
    unit vector toward the direction of the source and $\lambda$ is the
    wavelength.  Let $\ba_{i,k}$ have the same model as $\bAtt_{q,k}$
    with $\bbeta_i$ pointing towards the center of the $i$th pixel.  When a
    source is within a pixel but not exactly in the center we can model this
    mismatch as
\begin{align*}
\bat_{q,k}&=\frac{1}{\sqrt{P}}e^{\frac{j2\pi}{\lambda}\bXi^T\bQ_{k}(\bbeta_i+\bdelta_i)}\\
&= \ba_{i,k} \odot e^{\frac{j2\pi}{\lambda}\bXi^T\bQ_{k}\bdelta_i}\\
\end{align*}
    where $\bdelta_i=\tilde{\bbeta}_q-\bbeta_i$ and $i \in \MCF$.  Because
    both $\bbeta_i$ and $\tilde{\bbeta_q}$ are $3 \times 1$ unit vectors,
    each has only two degrees of freedom.  This means that we can
    parameterize the unknowns for the grid correcting problem using
    coefficients $\delta_{i,1}$ and $\delta_{i,2}$.  We will assume that
    when a source is added to the free set, its actual position is very
    close to the center of the pixel on which it was detected.  This means
    that $\delta_{i,1}$ and $\delta_{i,2}$ are within the pixel's width,
    denoted by $W$, and height, denoted by $H$.  In this case we can replace
    $\eqref{eq:subproblem}$ by a non-linear constrained optimization,
\begin{align}
\label{eq:doals}
\min_{\bdelta,\bsigma} &\frac{1}{2}\| \bb - \bAt(\bdelta)_{\MCF}\bsigma_{{\MCF}}\|_2^2 \notag \\
\text{s.t.}&-W/2<\delta_{i,1}<W/2\notag\\
&-H/2<\delta_{i,2}<H/2
\end{align}
    where $ \bAt(\bdelta)_{\MCF}$ contains only the columns corresponding to
    the set $\MCF$, $\bdelta_j$ is a vector obtained by stacking
    $\delta_{i,j}$ for $j=1,2$ and
\begin{equation}
    \bb=\brh-\bAt_{\MCU}\bsigma_{\MCU}.
\end{equation}
    This problem can also be seen as a direction of arrival (DOA) estimation
    which is an active research area and out of the scope of this paper.  A
    good review of DOA mismatch correction for MVDR beamformers can be found
    in \cite{chen2007}.

    Besides solving \eqref{eq:doals} instead of \eqref{eq:subproblem} in
    line 5 of the active set method we will also need to update the
    upper bounds and the standard deviations of the dirty images at the
    new pixel positions that are used in the other steps (e.g., line 3, 6
    and 13), the rest of the steps remain the same.  Because we have a
    good initial guess to where each source in the free set is, we
    propose a Newton based algorithm to do the correction.

%
%

\section{Implementation using Krylov Subspace based methods}
\label{sec:krylov}

From the active set methods  described in the previous section, we know that we need to solve \eqref{eq:subproblemLS} or \eqref{eq:subproblemMVDR} at each iteration. In this section we describe how to achieve this efficiently and without the need of storing the whole convolution matrix in memory.

\subsection{Lanczos algorithm and LSQR} 
When we are solving CLS or PWLS, we need to solve a problem of the form $\|\bb-\bM\bx\|_2^2$ as the first step of the active set iterations. For example, in \eqref{eq:subproblemLS} $\bM=\bAt_{\MCF}$. Note that it does not have to be a square matrix and usually it is ill-conditioned especially if the number of pixels is large. In general we can find a solution for this problem by first computing the singular value decomposition (SVD) of $\bM$  as
\begin{equation}
\label{eq:svd}
\bM=\bU\bS\bV^H,
\end{equation}
where $\bU$ and $\bV$ are unitary matrices and $\bS$ is a diagonal matrix with positive singular values. Then the solution $\bx$ to $\min \|\bb-\bM\bx\|^2$ is found by solving for $\by$ in
\begin{align}
\bS\by&=\bU^H\bb 
\end{align}
followed by setting
\begin{align}
\bx&=\bV\by.
\end{align}
Solving the LS problem with this method is expensive in both number of operations and memory usage, especially when the matrices $\bU$ and $\bV$ are not needed after finding the solution. As we will see shortly, looking at another matrix decomposition helps us to reduce these costs. For the rest of this section we use the notation given by \cite{paige1982}.

The first step in our approach for solving LS problem is to reduce $\bM$ to a lower bidiagonal form as follows
\begin{equation}
\bM=\bU\bB\bV^H,
\end{equation}
where $\bB$ is  a bidiagonal matrix  of the form 
\begin{equation}
\bB=\left[\begin{array}{cccc|c}
\alpha_1 & & & & \\
\beta_2&\alpha_2 &  & & \\
&\ddots&\ddots & & \\
& & \beta_r&\alpha_r &\\ \hline
& & & & \zeros\\
\end{array}\right],\\
\end{equation}
with $r=\rank(\bM)=\rank(\bB)$ and $\bU$,$\bV$ are unitary matrices (different than in \eqref{eq:svd}). This representation is not unique and without loss of generality we could choose $\bU$ to satisfy
\begin{equation}
\bU^H\bb=\beta_1\be_1
\end{equation}
where $\beta_1=\|\bb\|_2$ and $\be_1$ is a unit norm vector with its first element equal to one.

Using $\bB$, forward substitution gives the LS solution efficiently by solving $\by$ in 
\begin{equation}
\bB\by=\bU^H\bb=\beta_1\be_1 
\end{equation}
followed by
\begin{equation}
\bx=\bV\by\notag.
\end{equation}
 Using forward substitution we have
\begin{align}
y_1&=\frac{\beta_1}{\alpha_1}\\
\bx_1&=\bv_1 y_1,
\end{align}
followed by the recursion,
\begin{align}
y_{n+1}&=-\frac{\beta_{n+1}}{\alpha_{n+1}} y_{n}\\
\bx_{n+1}&=\bx_{n}+\bv_{n+1} y_{n+1}
\end{align}
for $n=1,\dots,M$ where $M < r$ is the iteration at which $\|\bM^H(\bM\bx_n-\bb)\|^2$ vanishes within the desired precision.
We can combine the bidiagonalization and solving for $\bx$ and avoid extra storage needed for saving $\bB$, $\bU$ and $\bV$. One such algorithm is based on a Krylov subspace method called the Lanczos algorithm \cite{golub1965}. We first initialize with

\begin{align}
\label{eq:initbeta}
\beta_1&=\|\bb\|_2\\
\label{eq:initu}
\bu_1&=\frac{\bb}{\beta_1}\\
\alpha_1&=\|\bM^H\bu_1\|_2\\
\bv_1&= \frac{\bM^H\bu_1}{\alpha_1}.
\end{align}
The iterations are then given by
\begin{align}
\label{eq:lanczos}
\begin{array}{l l}
    \beta_{n+1}&= \|\bM\bv_{n}-\alpha_{n}\bu_{n}\|_2\\
    \bu_{n+1}&=\frac{1}{\beta_{n+1}} (\bM\bv_n-\alpha_n\bu_n) \\
    \alpha_{n+1}&=\|\bM^H\bu_{n+1}-\beta_{n+1}\bv_{n}\|_2 \\
    \bv_{n+1}&=\frac{1}{\alpha_{n+1}} (\bM^H\bu_{n+1}-\beta_{n+1}\bv_n)
  \end{array}
\end{align}
for $ n=1,2,\dots,M$,
where $\bu_n^H\bu_n=\bv_n^H\bv_n=1$. This provides us with all the parameters needed to solve the problem.

However because of the finite precision errors, the columns of $\bU$ and $\bV$ found in this way loose their orthogonality as we proceed. In order to prevent this error propagation into the final solution $\bx$, different algorithms like Conjugate Gradient (CG), MINRES, LSQR, etc. have been proposed. The exact updates for $\bx_n$ and stopping criteria to find $M$ depends on the choice of algorithm used and therefor is not included in the iterations above.

An overview of Krylov subspace based methods, is given by \cite[pp.91]{choi2006}. This study shows that LSQR is a good candidate to solve LS problems when we are dealing with an ill-conditioned and non-square matrix. For this reason we will use LSQR to solve \eqref{eq:subproblemLS} or \eqref{eq:subproblemMVDR}. Because the remaining steps during the LSQR updates are a few scalar operations and do not have large impact on the computational complexity of the algorithm we will not go into the details.(see \cite{paige1982}) 

In the next section we discuss how to use the structure in $\bM$ to avoid storing the entire matrix in memory and how to parallelize the computations.  

%
%

\subsection{Implementation}

During the active set iteration we need to solve \eqref{eq:subproblemLS} and \eqref{eq:subproblemMVDR} where the matrix $\bM$ in LSQR is replaced by $\bAt_\MCF$ and $(\bR^{-T/2}\otimes \bR^{-1/2}) (\bAt\bD^{-1})_\MCF$ respectively. Because $\bAt$ has a Khatri-Rao structure and selecting and scaling a subset of columns does not change this, $\bAt_\MCF$ and $(\bAt\bD^{-1})_\MCF$ also have a Khatri-Rao structure. Here we will show how to use this structure to implement \eqref{eq:lanczos} in parallel and with less memory usage.

Note that the only time the matrix $\bM$ enters the algorithm is via the matrix-vector multiplications $\bM\bv_n$ and $\bM^H\bu_{n+1}$. As an example we will use $\bM=\bAt_{\MCF}$ for solving \eqref{eq:subproblemLS}.  
Let $\bk_n=\bAt_\MCF\bv_n$. We partition $\bk_n$ as $\bAt$ into
\begin{equation}
\bk_n=\begin{bmatrix}
\bk_{1,n}^T &\dots & \bk_{K,n}^T
\end{bmatrix}^T.
\end{equation}
Using the definition of $\bAt$ in \eqref{eq:bAt}, the operation $\bk_n=\bAt_\MCF\bv_n$ could also be performed using
\begin{equation}
\bK_{k,n}=\sum_{i \in \MCF} v_{i,n} \ba_{i,k}\ba_{i,k}^H.
\end{equation}
and subsequently setting 
\begin{equation}
\bk_{k,n}=\vect(\bK_{k,n}).
\end{equation}
This process can be highly parallelized because of the independence between the correlation matrices of each time snapshot. The matrix $\bK_{k,n}$ can then be used to find the updates in \eqref{eq:lanczos}.

The operation $\bM^H\bu$ in \eqref{eq:lanczos}, is implemented in a similar way. Using the beamforming approach (similar to Sec.\ref{power_constraint}), this operation can also be done in parallel for each pixel and each snapshot.

In both cases the calculations can be formulated as correlations and beamforming of parallel data paths which means that efficient hardware implementations are feasible. Also we can consider traditional LS or WLS solutions as a special case when all the pixels belong to the free set which means that those algorithms can also be implemented efficiently in hardware in the same way. Because during the calculations we work with a single beamformer at the time, the matrix $\bAt$ need not to be pre-calculated and stored in the memory. This makes it possible to apply image formation algorithms for large images when there is a memory shortage.

The computational complexity of the algorithm is dominated by the transformation between the visibility domain and image domain (correlation and beamforming). The dirty image formation and correlation have a complexity of $O(K p^2 I)$ this means that the worst case complexity of the active set algorithm is $O(T M K p^2 I)$ where $T$ is the number of active set iterations and $M$ is the maximum number of Krylov iterations. A direct implementation of CLEAN for solving the imaging problem presented in Sec.\ \ref{sec:imagingp} in similar way would have a complexity of $O(T K p^2 I)$. Hence the proposed algorithm is order $M$ times more complex, essentially because it recalculates the flux for all the pixels in the free-set while CLEAN only estimates the flux of newly added pixel. In practice many implementations of CLEAN use FFT instead of DFT (matched filter) for calculating the dirty image. Extending the proposed method to use similar techniques is possible and will be presented in future works.

\begin{figure*}[h!]
	\centering
	\includegraphics[width=0.85\textwidth]{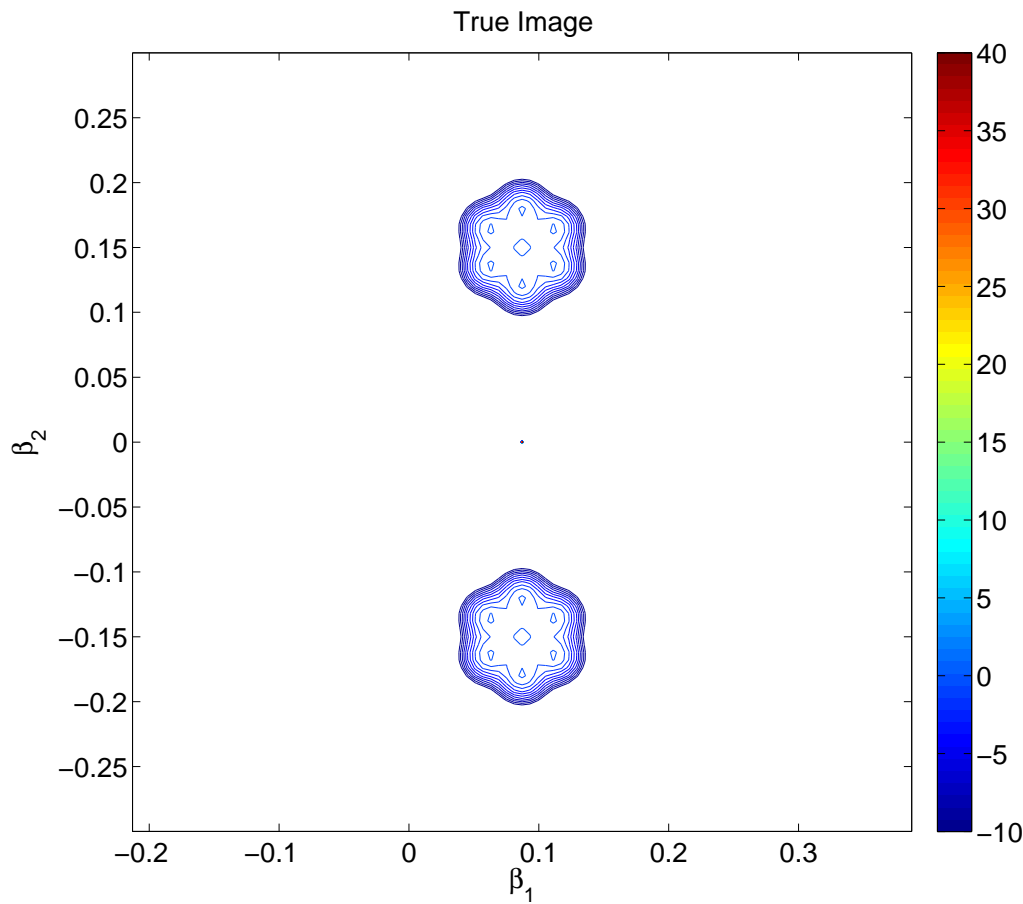}
	\caption{True source}
	\label{fig:true}
\end{figure*}

\begin{figure*}
    \subfloat[MF Dirty Image\label{fig:MF_Dirty}]{%
      \includegraphics[width=0.32\textwidth]{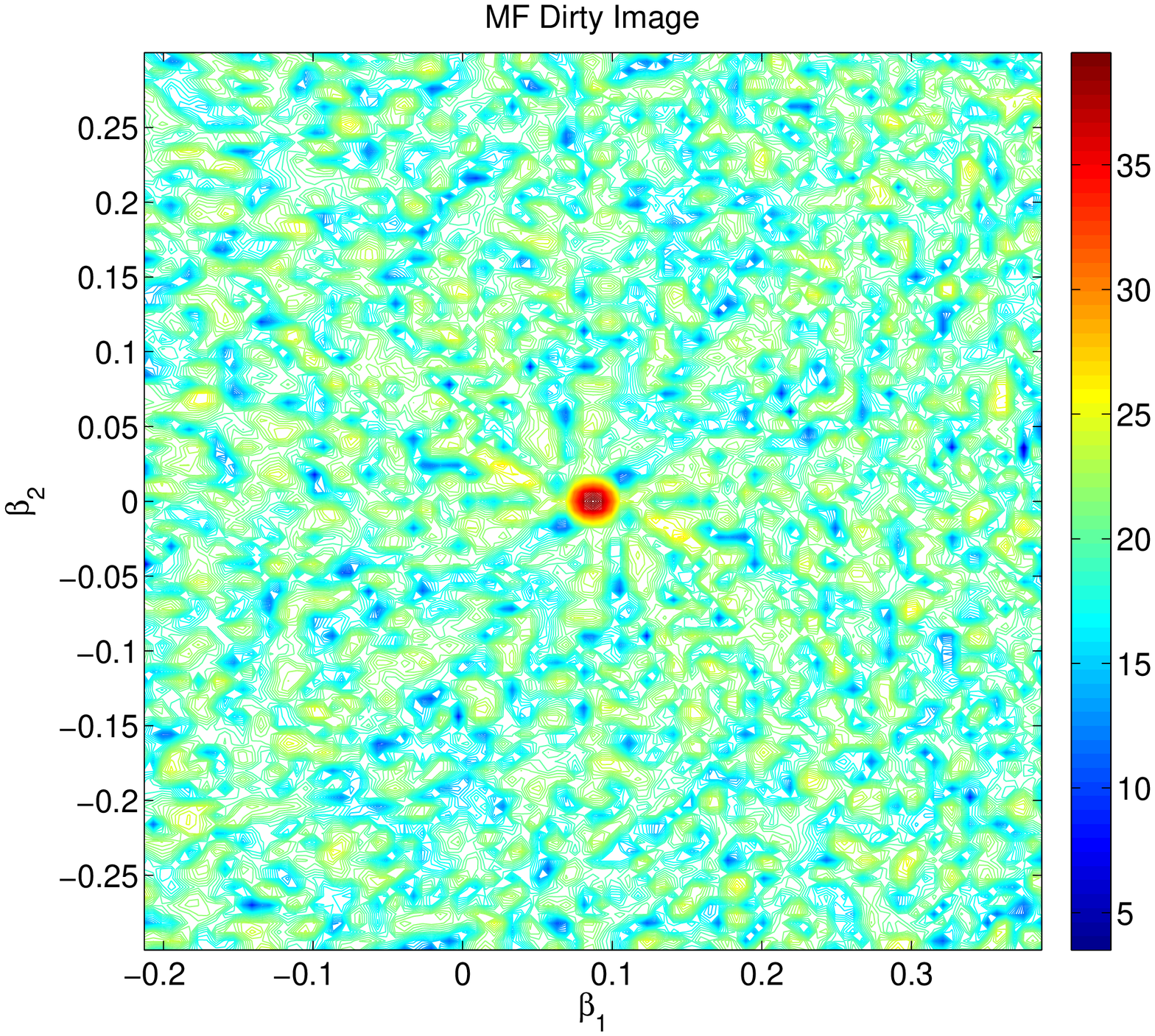}
    }
    \hfill
    \subfloat[Solution of the CLS image after convolution with a Gaussian beam\label{fig:LS_Clean_Gauss}]{%
      \includegraphics[width=0.32\textwidth]{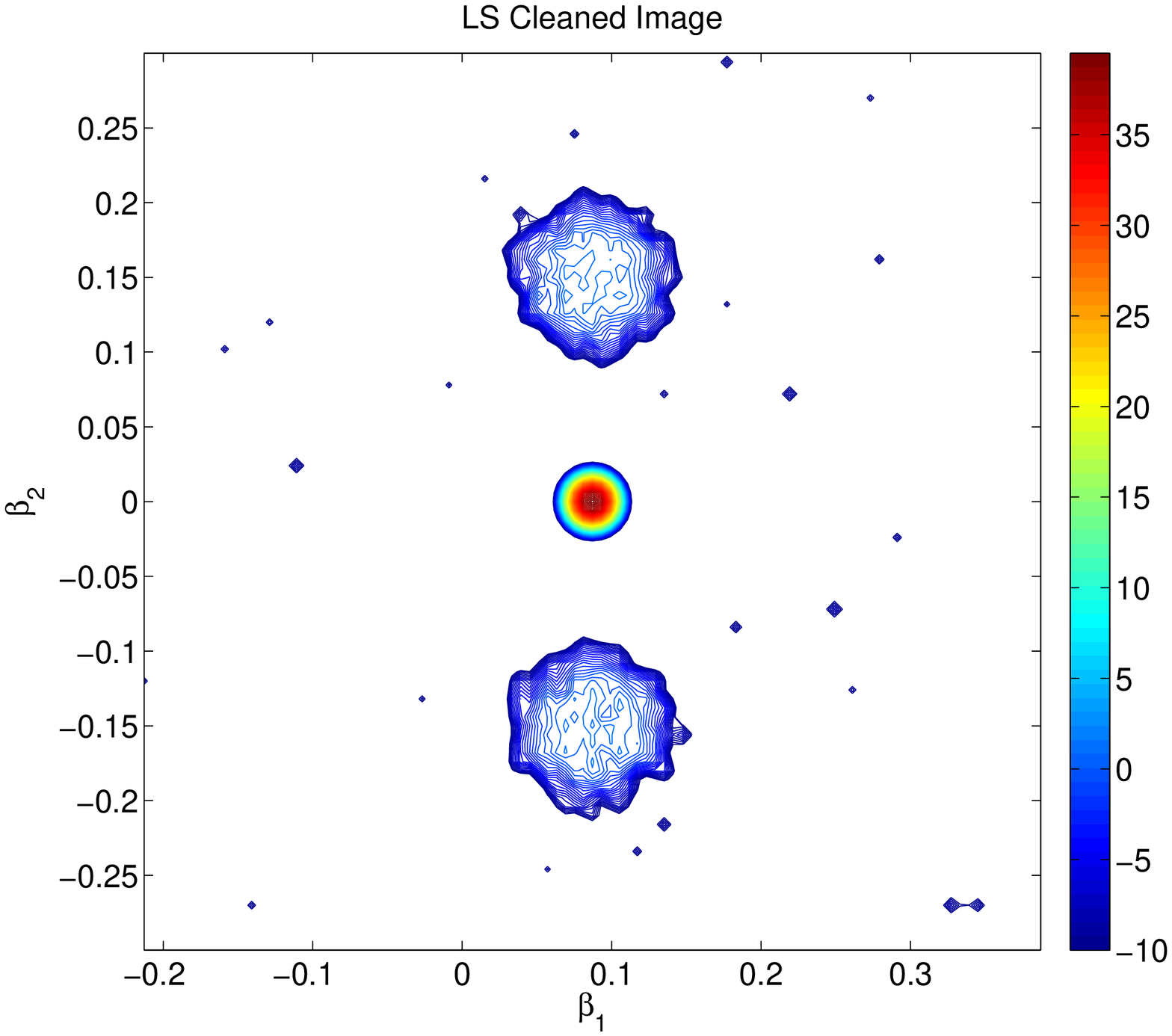}
    }
    \hfill
    \subfloat[CLS Image cross-section\label{fig:LS_cross}]{%
      \includegraphics[width=0.32\textwidth]{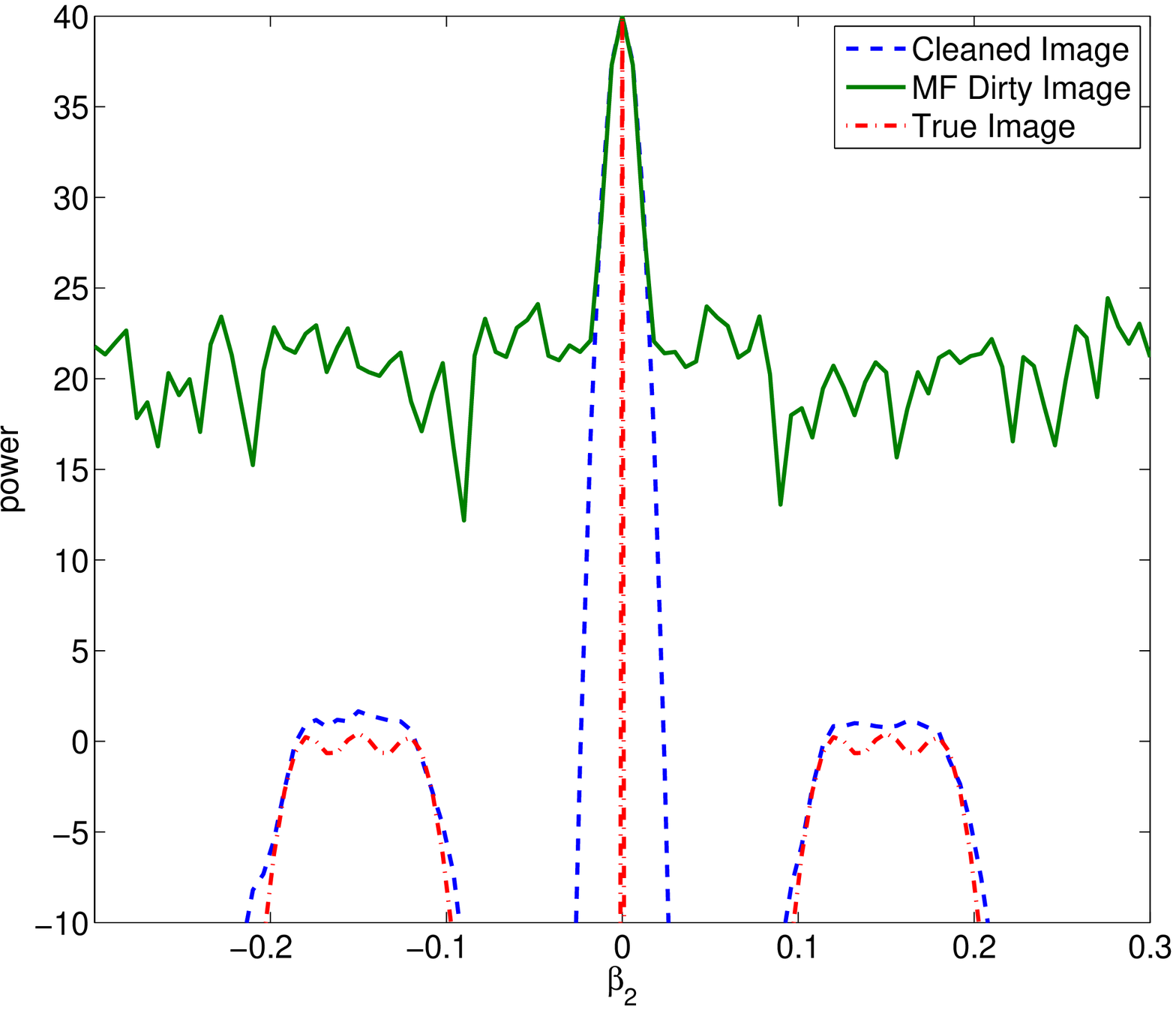}
    }
    \vfill
    \subfloat[MVDR dirty image\label{fig:MVDR_Dirty}]{%
      \includegraphics[width=0.32\textwidth]{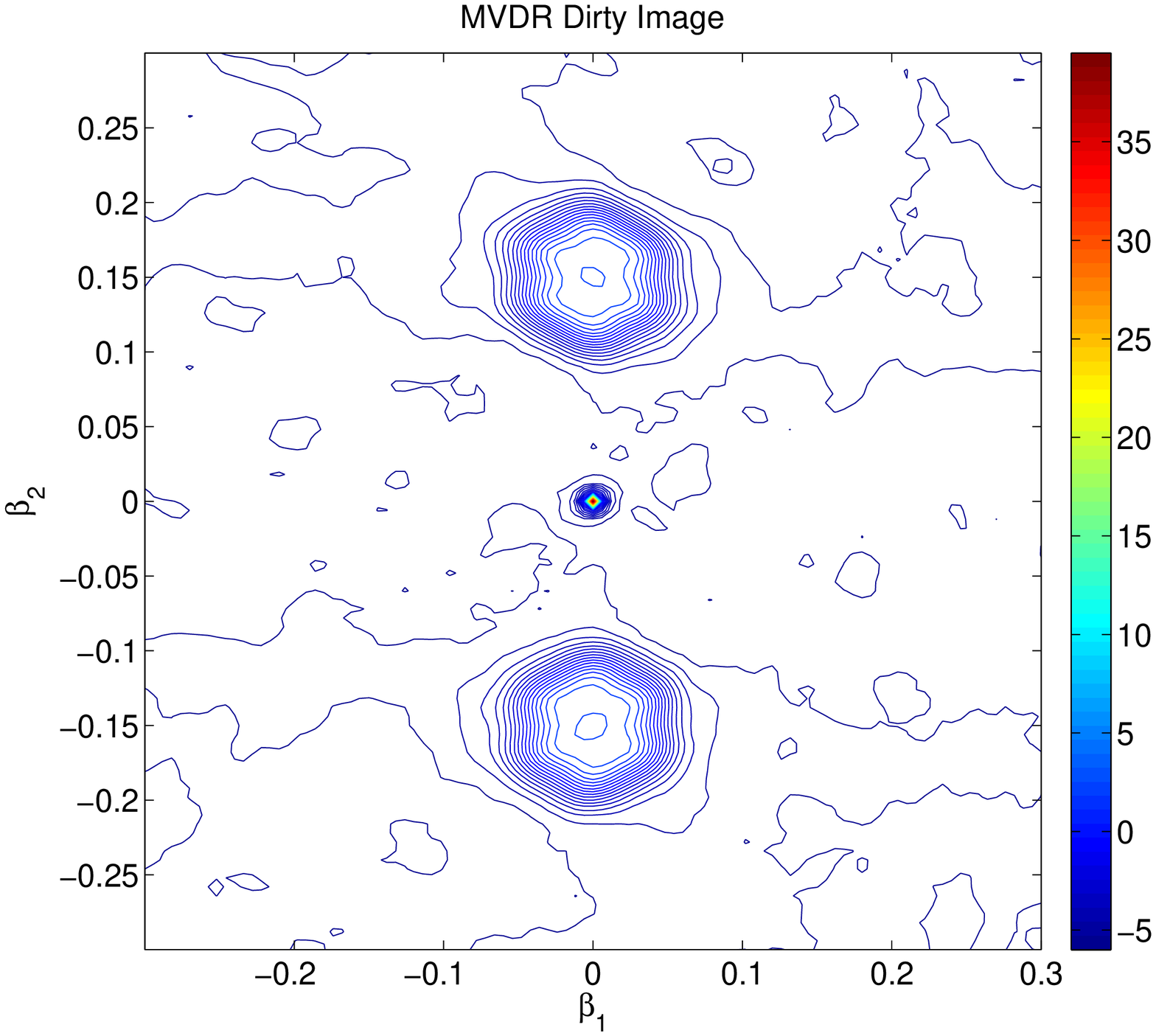}
    }
    \hfill
    \subfloat[Preconditioned WLS Image after convolution with a Gaussian beam\label{fig:MVDR_Clean_Gauss}]{%
      \includegraphics[width=0.32\textwidth]{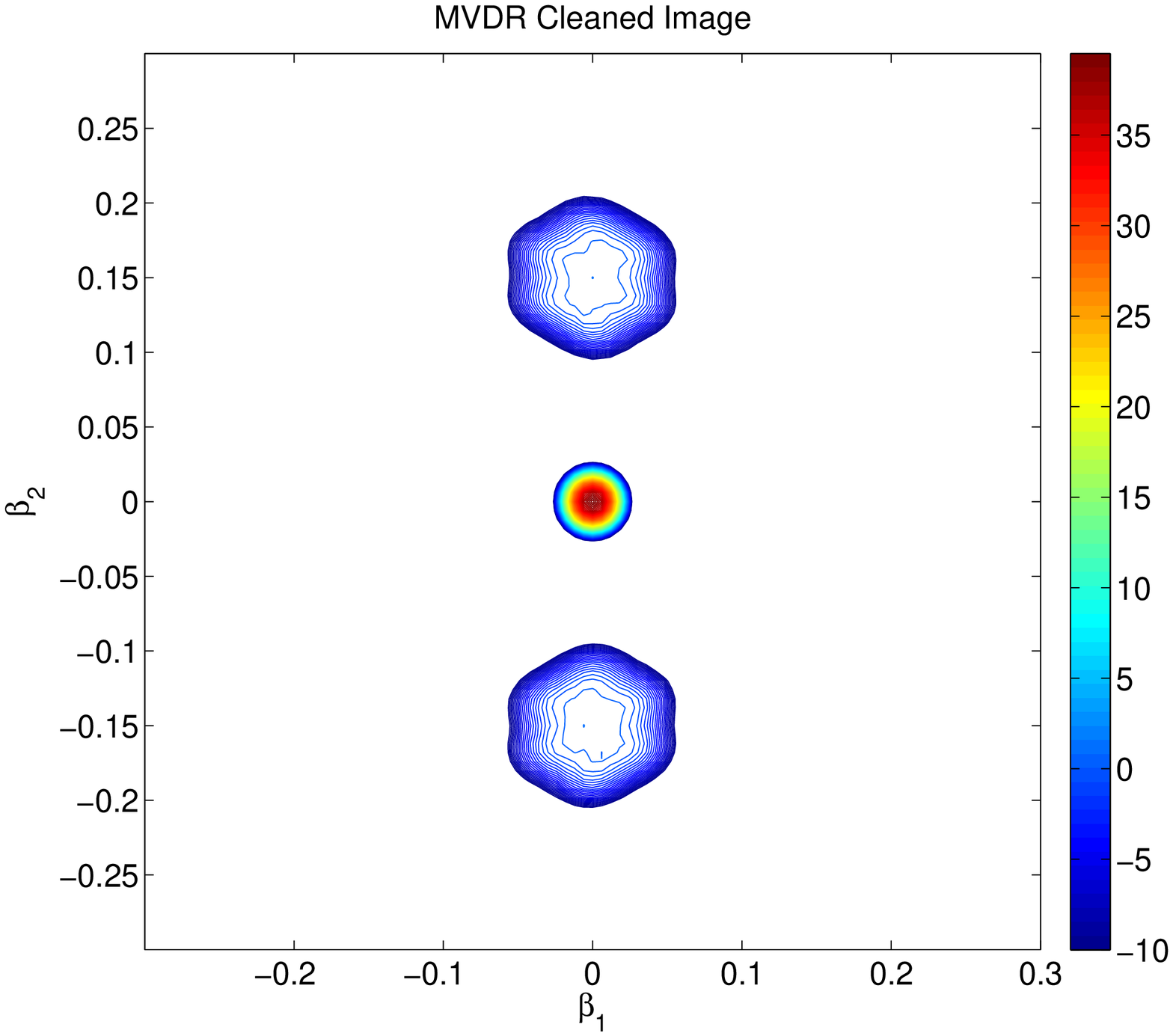}
    }
    \hfill
    \subfloat[MVDR dirty image and PWLS cross-section\label{fig:MVDR_cross}]{%
      \includegraphics[width=0.32\textwidth]{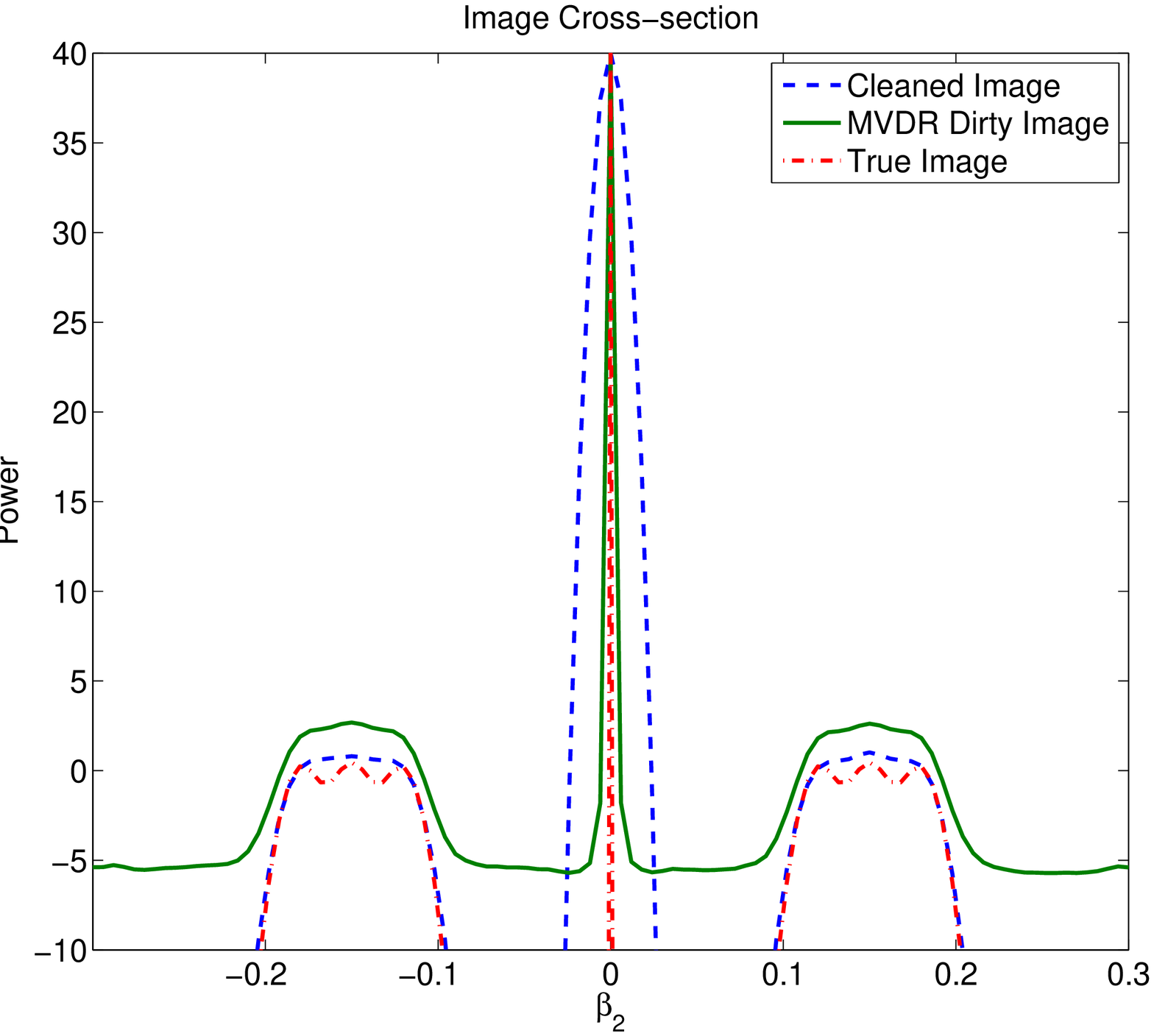}
    }

    \caption{Extended Source Simulations}
    \label{fig:superturb}
  \end{figure*}

\section{Simulations}
In this section we will evaluate the performance of the proposed method using simulations. Because the active set algorithm adds a single pixel to the free set at each step, it is important to investigate the effect of this procedure on extended sources and noise. For this purpose we will use a high dynamic range simulated image with a strong point source and two weaker extended sources in the first part of the simulations. In the second part we will make a full sky image using sources from the 3C catalog. We use the following definitions for the coordinate systems. A fixed coordinate system based on the right ascension ($\alpha$) and declination ($\delta$) of the sources
\[
\bbeta=\begin{bmatrix}
\cos(\delta)\cos(\alpha) \\
\cos(\delta)\sin(\alpha) \\
\sin(\delta).
\end{bmatrix}
\]
The corresponding $(l,m,n)$ coordinates $\bs$ that take earth rotation into account are given by
\[
\bs=\bQ_k(L,B)\bbeta.
\]

\subsection{Extended Sources}

An array of 100 dipoles ($p=100$) with random distribution is used with the frequency range of 58-90 MHz from which we will simulate three equally spaced channels. Each channel has a bandwidth of $195$ kHz and is sampled at Nyquist-rate. These specification have been chosen the same as for LOFAR telescope in LBA modes \cite{refIdLOFAR}. LOFAR uses $1$ second snapshots and we will simulate using only two snapshots, this means that  $K=2$.
The simulated source is a combination of a strong point source and two extended structures. The extended sources are composed from seven Gaussian shaped sources, one in the middle and 6 on a hexagon around it. Figure \ref{fig:true} shows the simulated image in dB scale. The background noise level that is added is at $-10$ dB which is also $10$ dB below the the extended sources.

Figures \ref{fig:MF_Dirty} and \ref{fig:MVDR_Dirty} show the matched filter and MVDR dirty images respectively. Figures \ref{fig:LS_Clean_Gauss} and \ref{fig:MVDR_Clean_Gauss} show the reconstructed images, after deconvolution and smoothing with a Gaussian clean beam, for the CLS and PWLS deconvolution with MF and MVDR dirty images as upper bounds respectively. A cross section of the images has been illustrated in Figures \ref{fig:LS_cross} and \ref{fig:MVDR_cross}.
Remarks are:
\begin{itemize}
\item  As expected the MVDR dirty image has a much better dynamic range and lower side-lobes;
\item Due to a better initial dirty image and upper bound the preconditioned WLS deconvolution gives a better cleaned image. However a trade-off is made between the resolution of the point source and the correct shape of the extended sources when we use the Gaussian beam to smoothen the image.
\item The cross sections show the accuracy of the magnitudes. This shows that not only the shape but also the magnitude of the sources are better estimated using PWLS.
\end{itemize}

\subsection{Full Sky with 3C Sources}
In this part we describe a simulation for making an all sky image. The array setup is the same as before with the same number of channels and snapshots. A background noise level of $0$ dB (with respect to 1 Jansky) is added to the sky.

In this simulation we check which sources from the 3C catalog are visible
at the simulated date and time. From these we have chosen 20 sources that
represent the magnitude distribution on the sky and produce the highest
dynamic range available in this catalog. Table \ref{table:sources} shows
the simulated sources with corresponding parameters. The coordinates are
the $(l,m)$ coordinates at the first snapshot. Because the sources are not
necessarily on the grid point we have chosen to do the active set
deconvolution in combination with grid correction on the free set as
described in Sec.\ \ref{sec:strong_sources}.

Figures \ref{fig:3C_positions} and \ref{fig:3C_powers} show the position and power estimates for the sources that are detected during the deconvolution process. Figure \ref{fig:3C_MF_Dirty} shows the full sky MF dirty image. The contoured version of the reconstructed image with minimum contour $3$ dB above the noise level is shown in Figure \ref{fig:3C_contour} and the final reconstructed image with the residual added to it is give in Figure \ref{fig:3C_Final}

Remarks:
\begin{itemize}
\item  The algorithm stops after adding the correct number of sources based on the detection mechanism we have incorporated in the active set method;
\item Because of the grid correction no additional sources are added to compensate for incorrect power estimates on the grids;
\item All 20 sources are visible in the final reconstructed image and no visible artifacts are added to image.
\end{itemize}

\begin{table}
\caption{Simulated Sources from the 3C Catalog}
\label{table:sources}
\centering
\begin{tabular}{|l|c|c|c|}
\hline
Names & $l$ & $m$ & Flux \\
\hline
3C   461 & -0.30485 & 0.19131 &11000 \\
3C   134 &  0.59704 &-0.02604 &   66 \\
3C   219 &  0.63907 &  0.6598 &   44 \\
3C  83.1 &  0.28778 &-0.13305 &   28 \\
3C    75 &  0.30267 &  -0.684 &   23 \\
3C    47 &-0.042882 &-0.51909 &   20 \\
3C 399.2 & -0.97535 & 0.20927 &   19 \\
3C   6.1 &-0.070388 & 0.47098 &   16 \\
3C   105 &  0.57458 &-0.60492 &   15 \\
3C   158 &   0.9017 &-0.12339 &   14 \\
3C   231 &  0.28956 & 0.72005 &   13 \\
3C   303 &  -0.1511 & 0.95402 & 12.5 \\
3C 277.1 &  0.12621 & 0.93253 &   12 \\
3C   320 &  -0.3597 & 0.93295 & 11.5 \\
3C 280.1 &  0.15171 & 0.98709 &   11 \\
3C 454.2 & -0.29281 & 0.31322 & 10.5 \\
3C   458 & -0.61955 &-0.56001 &   10 \\
3C 223.1 &  0.67364 & 0.68376 &  9.5 \\
3C    19 & -0.23832 &-0.30028 &    9 \\
3C 437.1 & -0.83232 &-0.24924 &    5 \\
\hline
\end{tabular}

\end{table}

\begin{figure*}
    \subfloat[Location Estimates\label{fig:3C_positions}]{%
      \includegraphics[width=0.47\textwidth]{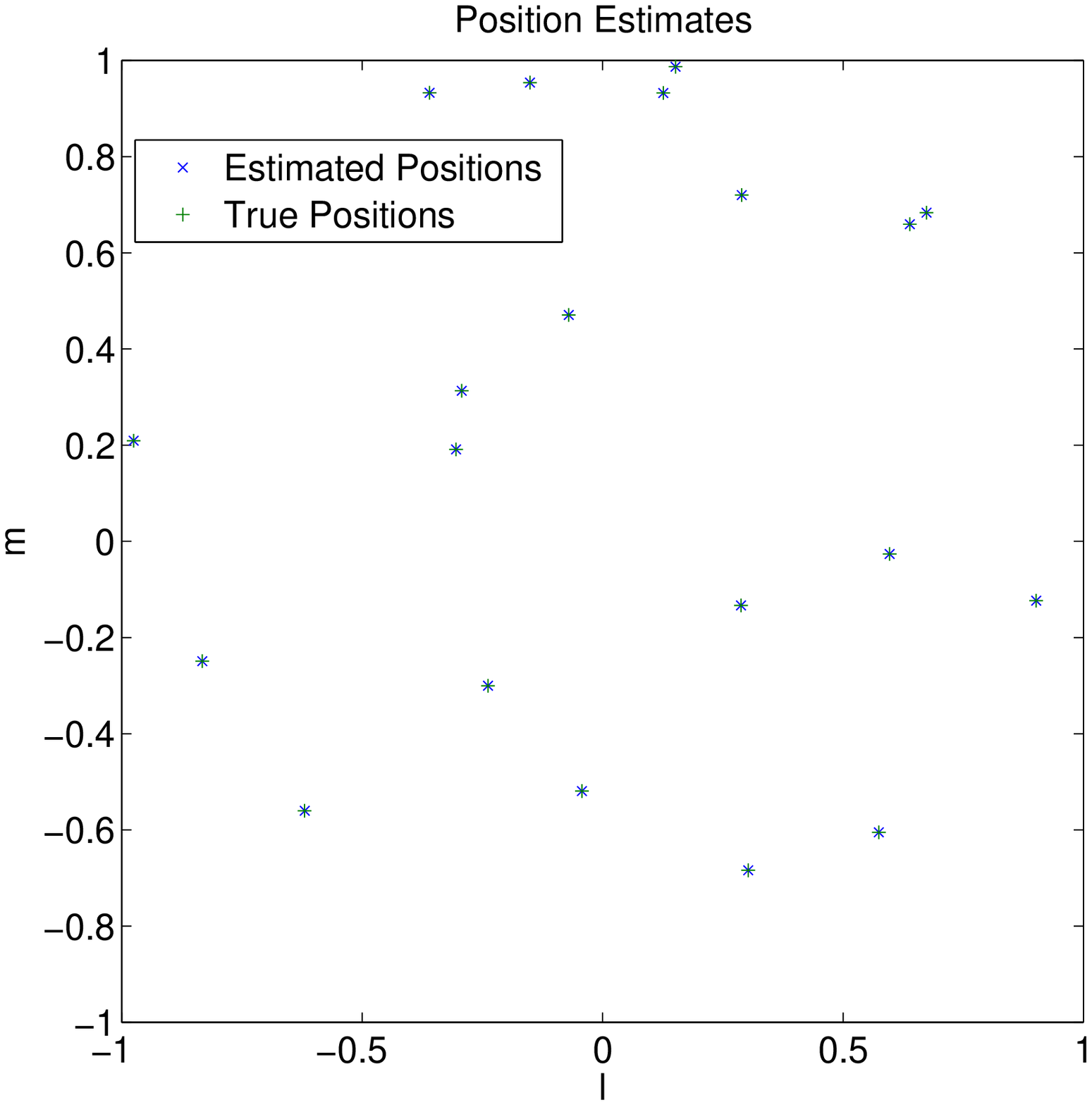}
    }
    \hfill
    \subfloat[Flux Estimates\label{fig:3C_powers}]{%
      \includegraphics[width=0.47\textwidth]{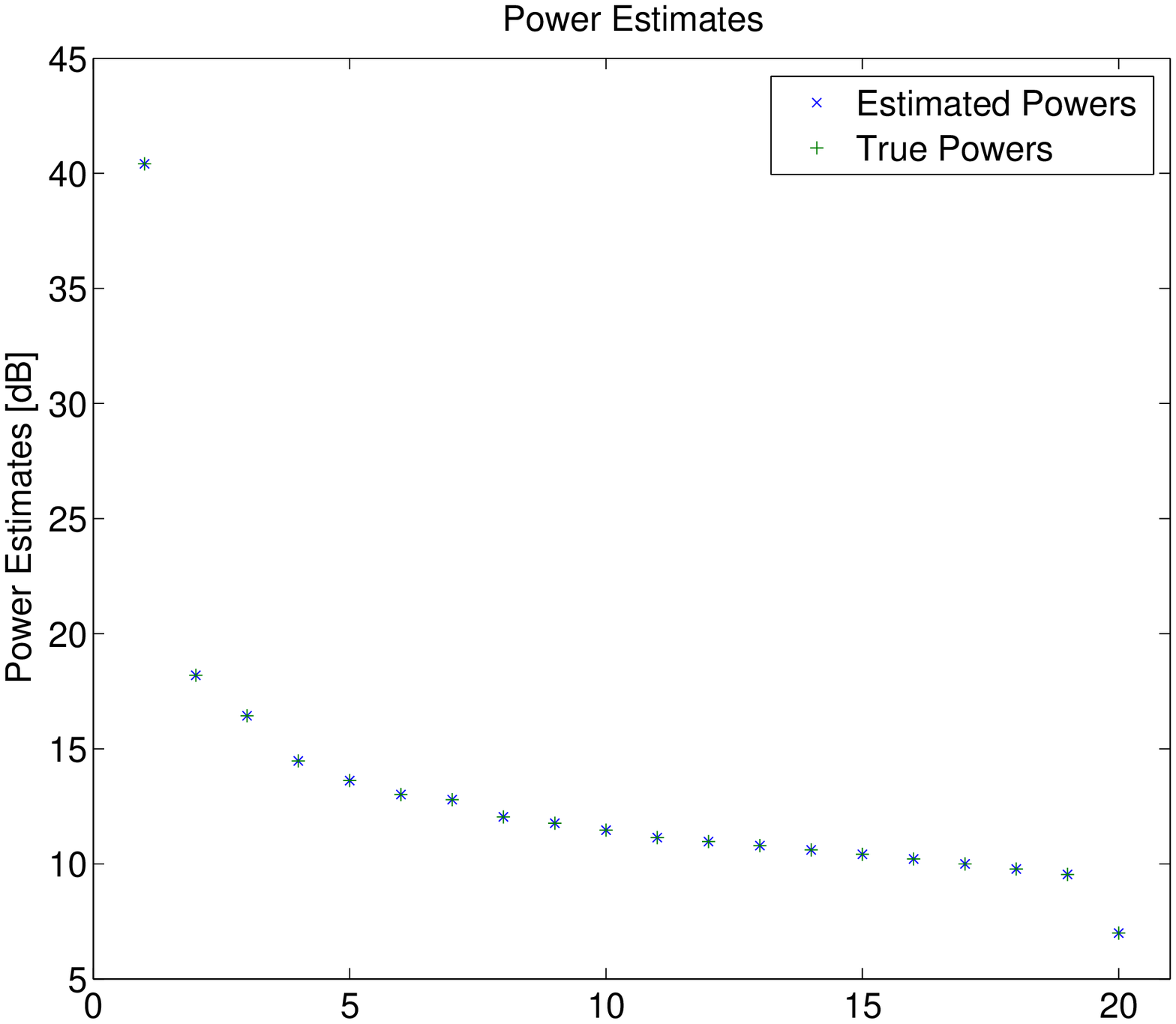}
    }
    \vfill
    \subfloat[Full Sky MF Dirty Image \label{fig:3C_MF_Dirty}]{%
      \includegraphics[width=0.47\textwidth]{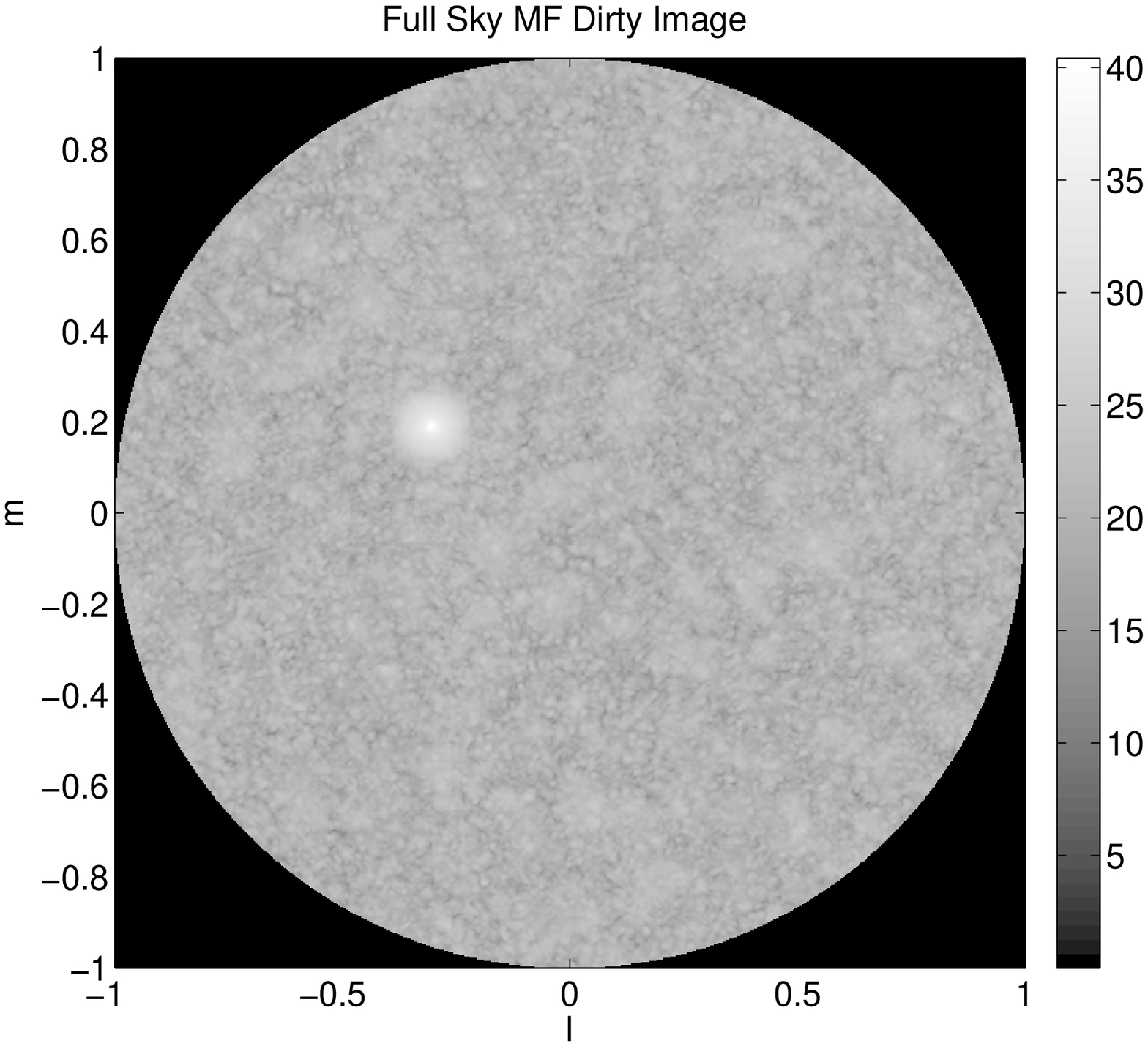}
    }
    \hfill
    \subfloat[Reconstructed Image-scale\label{fig:3C_contour}]{%
      \includegraphics[width=0.47\textwidth]{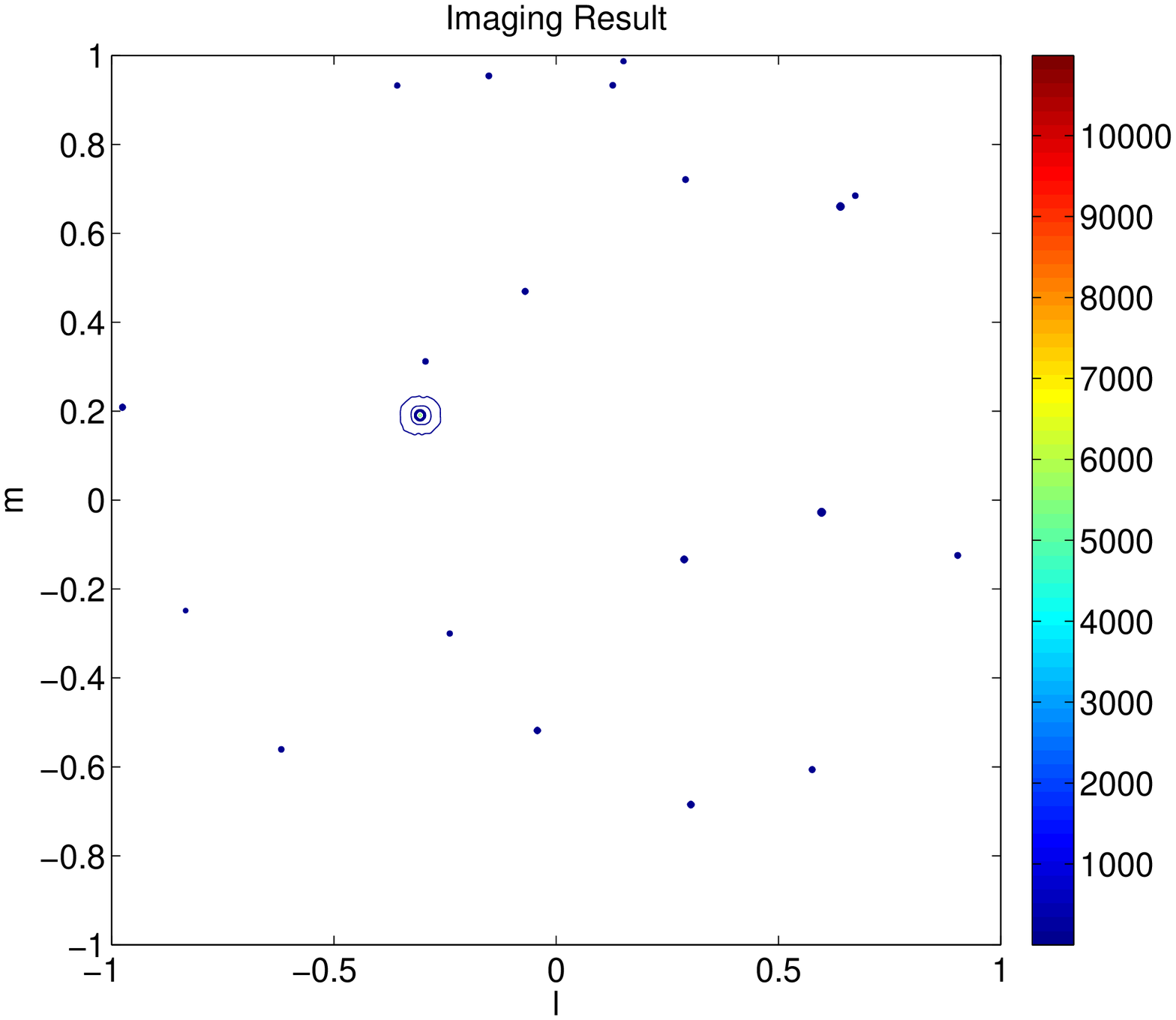}
    }

    \caption{Point Source Simulations}
    \label{fig:superturb1}
  \end{figure*}

\begin{figure*}
	\centering
	\subfloat{
	\includegraphics[width=0.85\textwidth]{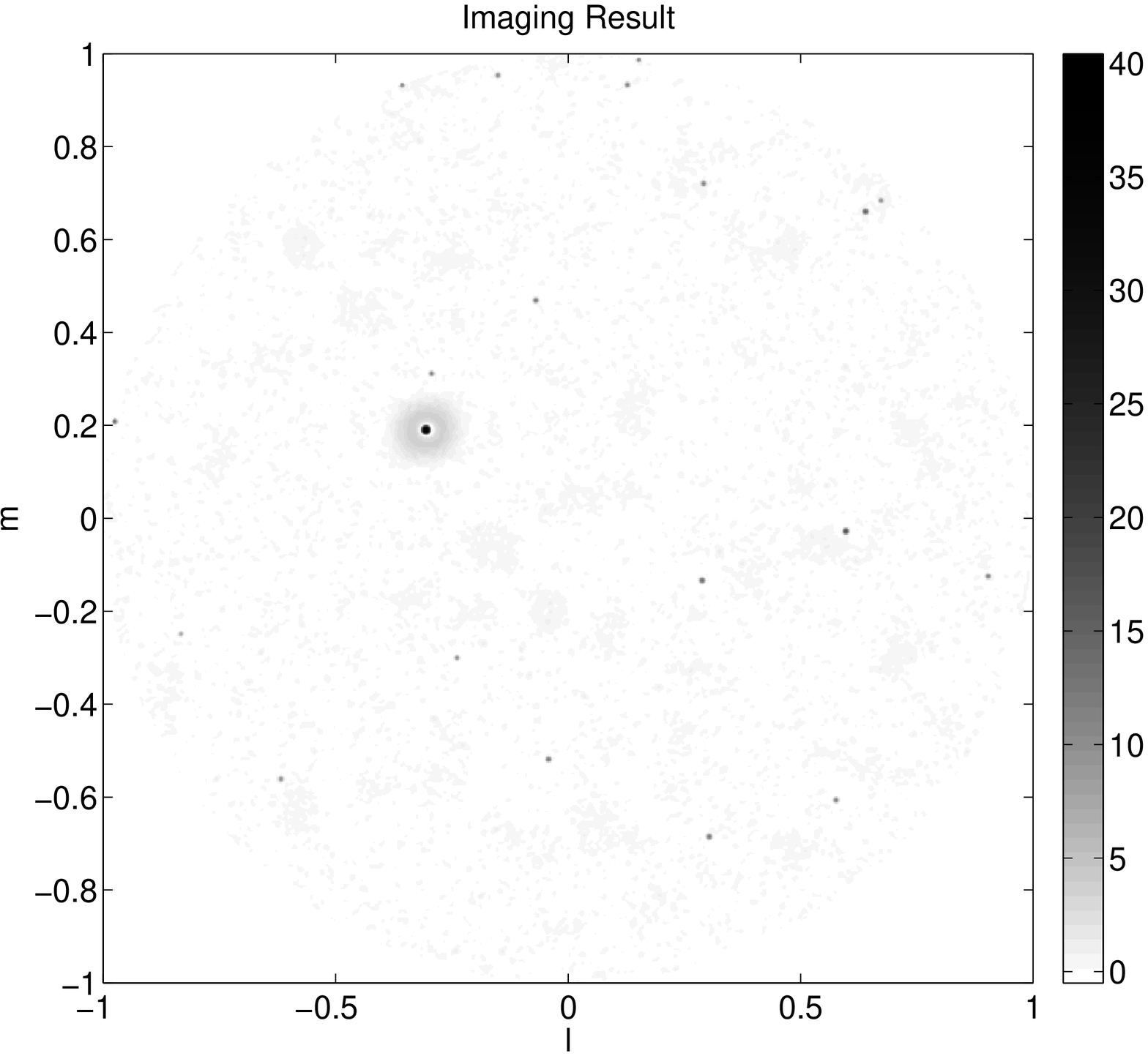}}
	\caption{Reconstructed Image plus the Residual Image}
	\label{fig:3C_Final}
\end{figure*}

\section{Conclusions}
Based on a parametric model and power constraints, we have formulated image deconvolution as an optimization problem with inequality constraints which we have solved using an active set based method. The relation between the proposed method and sequential source removing techniques is explained. The theoretical background of the active set methods can be used to gain better insight into how the sequential techniques work.

The Khatri-Rao structure of the data model is used in combination with Krylov based techniques to solve the linear systems involved in the deconvolution process with less storage and complexity.  We have introduced a preconditioned WLS cost function with a gradient that is related to the MVDR dirty image. Using simulation we have shown that the solution to the preconditioned WLS has improved spatial structure and improved power estimates. 

In this paper we have discussed the bidiagonalization and Krylov approach to solve the system of linear equations. The main reason for this is to reduce the storage needed for the deconvolution matrix. It is easy to verify that the active set updates can be translated into rank one updates and downdates of the deconvolution matrix. There are other matrix decompositions like QR decomposition that can take advantage of this fact. Knowing that the Khatri-Rao structure of the matrix does not change by adding or removing columns, it is interesting for future works to investigate whether rank one changes can be combined with the Krylov based techniques.

%
%
\appendices
\section{Upper Bounds on Image Powers}
\label{appedix:1}

 To find the confidence intervals for the dirty images we need to find estimates for the variance of both matched filter and MVDR dirty images. In our problem the sample covariance matrix is obtained by squaring samples from a Gaussian process. This means that $N\bRh \sim \MCW_p(\bR,N)$ where $\MCW_p(\bR,N)$ is the Wishart distribution function of order $p$ with expected value equal to $\bR$ and $N$ degrees of freedom. For any deterministic vector $\bzeta$,
\begin{equation}
N\bzeta^H\bRh\bzeta \sim \bzeta^H\bR\bzeta~ \chi^2(N).
\end{equation}
where $\chi^2(N)$ is the standard $\chi^2$ distribution with $N$ degrees of freedom.
In radio astronomical applications $N$ is usually very large and we can approximate this $\chi^2$ distribution with a Gaussian such that $\bzeta^H\bRh\bzeta \sim \MCN(\bzeta^H\bR\bzeta,(\bzeta^H\bR\bzeta)^2/N)$. The variance of the matched filter dirty image is given by
\begin{equation*}
\Var(\MFDIind{i})=\frac{1}{NK^2}\sum_k(\ba_{i,k}^H\bR\ba_{i,k})^2
\end{equation*}
Using this result we can find the $x\%$ confidence interval which results in an increase of the upper bound such that
\begin{equation}
\bsigma \leq \SMFDI + \alpha \sqrt{\Var(\SMFDI)}
\end{equation}
where $\alpha$ is chosen depending on $x$. Requiring at most a single false detection on the entire image translate into $\alpha \approx 6$.

When we estimate the MVDR dirty image from sample covariance matrices we need to be more careful, mainly because the result is biased and we need to correct for that bias. For each pixel of the MVDR dirty image obtained from sample covariance matrices we have
\begin{align}
\UMVDRDIind{i}=K g(Z)=\frac{K}{\sum_k \ba_{i,k}^H\bRh_{k}^{-1}\ba_{i,k}}\notag \\
\end{align}
where $g(Z)=1/Z$ and $Z=\sum_k \ba_{i,k}^H\bRh_{k}^{-1}\ba_{i,k}$. Using a perturbation model $Z=Z_0+\Delta Z$ and a Taylor approximation we find
\begin{align}
\label{eq:gz}
g(Z)&\approx\frac{1}{Z_0}-\frac{1}{Z_0^2}\Delta Z\notag \\
&\approx \frac{1}{Z_0^2}(Z_0-\Delta Z).
\end{align}
Let $Z_0=\MCE\{Z\}$ then $\MCE\{\Delta Z\}=0$ and $\MCE\{g(Z)\}\approx1/Z_0$. We would like this estimate to be unbiased which means that we want
\begin{equation}
\MCE\{g(Z)\}\approx\frac{1}{\sum_k \ba_{i,k}^H\bR_{k}^{-1}\ba_{i,k}}
\end{equation}
however we have,
\begin{align}
Z_0&=\sum_k \ba_{i,k}\MCE\{\bRh_{k}^{-1}\}\ba_{i,k} \notag \\
&=\sum_k \ba_{i,k}^H\frac{N\bR_{k}^{-1}}{N-p}\ba_{i,k} \notag \\
&=\frac{N}{N-p}\sum_k \ba_{i,k}^H\bR_{k}^{-1}\ba_{i,k}
\end{align}
where we have used $\MCE\{\bRh^{-1}\}=\frac{N}{N-p}\bR^{-1}$ \cite{shaman1980}. So in order to remove this bias we need to scale it by a correction factor
\begin{equation}
\label{eq:C}
C=\frac{N}{N-p}
\end{equation}
and
\begin{equation}
\UMVDRDIind{i}=CK g(Z).
\end{equation}

Now we need to find an estimate for the variance of the MVDR dirty image.  Using \eqref{eq:gz} we see that the first order approximation of $\Var(g(Z))\approx\Var(Z)/Z_0^4$. We find $\Var(Z)$ using the independence of each snapshot so we can write
\begin{equation}
\Var(Z)=\sum_k \Var(\ba_{i,k}^H\bRh_{k}^{-1}\ba_{i,k}).
\end{equation}
In order to find $\Var(\ba_{i,k}^H\bRh_{k}^{-1}\ba_{i,k})$ we need to use some properties of the complex inverse Wishart distribution. A matrix has complex inverse Wishart distribution if it's inverse has a complex Wishart distribution \cite{shaman1980}. Let us define an invertible matrix $\bB$ as
\begin{equation}
\bB=\begin{bmatrix}\ba_{i,k} & \bB_1 \end{bmatrix}
\end{equation}
then $\bX=(\bB\bRh^{-1}\bB^H)/N$ has an inverse Wishart distribution because $\bX^{-1}=N(\bB^{-H}\bRh\bB^{-1})$ has a Wishart distribution. In this case $\bX_{11}=(\ba_{i,k}^H\bRh^{-1}\ba_{i,k})/N$ also has an inverse Wishart distribution with less degrees of freedom. The covariance of an inverse Wishart matrix is derived in \cite{shaman1980}, however because we are dealing only with one element, this results simplifies to
\begin{equation}
\Var(N\bX_{11})=\frac{N^2}{(N-p)^2(N-p-1)}(\ba_{i,k}^H\bR^{-1}\ba_{i,k})^2.
\end{equation}
The variance of the unbiased MVDR dirty image is thus given by
\begin{align}
\Var(\UMVDRDIind{i})&=\Var(C K g(Z))\notag \\
&\approx \frac{K^2}{(N-p-1)}\frac{\sum_k (\ba_{i,k}^H\bR_{k}^{-1}\ba_{i,k})^2}{\left(\sum_k \ba_{i,k}\bR_{k}^{-1}\ba_{i,k}\right)^4}.\notag
\end{align}
Now that we have the variance we can use the same method that we used for MF dirty image to find $\alpha$ and
\begin{equation}
\bsigma \leq \UMVDRDI + \alpha \sqrt{\Var(\UMVDRDI)}
\end{equation}

%
%

\section{Optimum Beamformer}
\label{appen:assc}

We have already defined the problem of finding the beamformer for optimum upper bound as
\begin{align}
\bw_{i,\text{opt}}&=\arg \min_{\bw} \bw^H \bR \bw \\
& \text{s.t.} \bw^H(\bI_K \circ \bA_i)(\bI_K \circ \bA_i)^H\bw=1 \notag
\end{align}
Following standard optimization techniques we define the Lagrangian and take derivatives with respect to $\bw$ and the Lagrange multiplier $\mu$ and we find
\begin{align}
\label{ew:grad}
\bw&=\mu \bR^{-1}(\bI_K \circ \bA_i)(\bI_K \circ \bA_i)^H\bw \\
\label{ew:equality}
1&=\bw^H(\bI_K \circ \bA_i)(\bI_K \circ \bA_i)^H\bw
\end{align}
 Because $\bR$ is full--rank and \eqref{ew:equality} we can model $\bw$ as
\begin{equation}
\bw=\mu \bR^{-1}(\bI_K \circ \bA_i) \bx.
\end{equation}
Filling back into \eqref{ew:grad} we have
\begin{equation}
\begin{array}{l}
\mu \bR^{-1}(\bI_K \circ \bA_i) \bx \\=\mu^2 \bR^{-1}(\bI_K \circ \bA_i)(\bI_K \circ \bA_i)^H\bR^{-1}(\bI_K \circ \bA_i) \bx 
\end{array}
\end{equation}
and 
\begin{equation}
\begin{array}{l}
(\bI_K \circ \bA_i) \bx\\=\mu (\bI_K \circ \bA_i)(\bI_K \circ \bA_i)^H\bR^{-1}(\bI_K \circ \bA_i) \bx
\end{array}
\end{equation}
multiplying both sides by $(\bI_K \circ \bA_i)^H$ we get
\begin{equation}
\label{eq:xgrad}
\bx=\mu (\bI_K \circ \bA_i)^H\bR^{-1}(\bI_K \circ \bA_i) \bx.
\end{equation}
Doing the same for \eqref{ew:equality} we have
\begin{equation}
\begin{array}{l}
\mu^2 \bx^H(\bI_K \circ \bA_i)^H \bR^{-1}(\bI_K \circ \bA_i)(\bI_K \circ \bA_i)^H \bR^{-1}(\bI_K \circ \bA_i) \bx \\
=1.
\end{array}
\end{equation}
Now we use \eqref{eq:xgrad} and we find
\begin{equation}
\bx^H\bx=1
\end{equation}
which makes finding $\bx$ an eigenvalue problem. By taking a closer look at the matrix $ (\bI_K \circ \bA_i)^H\bR^{-1}(\bI_K \circ \bA_i)$ we find that this matrix is diagonal
\begin{equation}
\begin{array}{l}
 (\bI_K \circ \bA_i)^H\bR^{-1}(\bI_K \circ \bA_i)\\=
\begin{bmatrix}
\ba_{i,1}^H\bR_1^{-1}\ba_{i,1} & \zeros & \dots & \zeros \\
\zeros & \ba_{i,2}^H\bR_2^{-1}\ba_{i,2} &&\vdots \\
\vdots &  &\ddots & \zeros\\
\zeros &\dots &\zeros & \ba_{i,K}^H\bR_K^{-1}\ba_{i,K}
\end{bmatrix}
\end{array}
\end{equation}
and hence $\bx=\be_m$ is an elementary vector with all entries equal to zero except for $m$th entry which equals unity. $m$ is the index corresponding to largest eigenvalue, $\lambda_{\text{max}}$, and from \eqref{eq:xgrad} we have $\mu=1/\lambda_{\text{max}}$. Filling back for $\bw$ we find
\begin{equation}
\bw_{i,\text{opt}}=\frac{1}{\ba_{i,m}\bR_m^{-1}\ba_{i,m}}\bR^{-1}(\be_m \otimes \ba_{i,m})
\end{equation}
and the output of the beamformer 
\begin{align}
\sigma_{opt}&= \bw_{i,\text{opt}}^H\bR\bw_{i,\text{opt}} \notag \\
&=\frac{\ba_{i,m}^H\bR^{-1}_m\ba_{i,m}}{(\ba_{i,m}^H\bR^{- 1}_m\ba_{i,m})^2} \notag \\
&=\frac{1}{\ba_{i,m}^H\bR_m^{-1}\ba_{i,m}} \notag \\
&=\min_k\left(\frac{1}{\ba_{i,k}^H\bR_k^{-1}\ba_{i,k}}\right)
\end{align}

\bibliographystyle{IEEEtran}
\bibliography{biblio,amir}
\end{document}